\documentclass[%
 reprint,
 preprintnumbers,
nofootinbib,
 amsmath,amssymb,
 aps,
prd
]{revtex4-1}

\usepackage{graphicx}
\usepackage{dcolumn}
\usepackage{bm}

\DeclareMathOperator{\MeV}{MeV}
\DeclareMathOperator{\GeV}{GeV}
\DeclareMathOperator{\TeV}{TeV}

\begin{document}

\preprint{CYCU-HEP-15-12}


\title{
\boldmath $Z^\prime$-induced FCNC decays of top, beauty and strange quarks
}

\author{Kaori Fuyuto$^{a}$, Wei-Shu Hou$^{b}$, and Masaya Kohda$^{c}$}
 \affiliation{
$^{a}$Department of Physics, Nagoya University, Nagoya 464-8602, Japan\\
$^{b}$Department of Physics, National Taiwan University, Taipei 10617, Taiwan\\
$^{c}$Department of Physics, Chung-Yuan Christian University, Chung-Li 32023, Taiwan
}


\begin{abstract}
Anomalous $b\to s$ transitions from LHCb data may suggest a new massive gauge
boson $Z'$ that couples to the left-handed $b\to s$ current, 
which in turn implies a coupling to the $t\to c$ current.
In this paper, we study flavor-changing neutral current (FCNC) decays
of the top quark induced by a $Z'$ boson, namely $t\to c Z'$, 
based on a model of the gauged $L_\mu - L_\tau$ symmetry
 (the difference between the muon and tauon numbers) 
with vector-like quarks, which was introduced to explain the anomalous LHCb data.
We illustrate that searching for $t\to cZ'$ via $Z'\to \mu^+\mu^-$ 
with LHC Run 1 data can already probe a parameter region which is 
unexplored by $B$ physics for the $Z'$ mass around ${\cal O}(10)$ GeV or more.
We further extend the model to very light $Z'$ with mass below 400 MeV,
which is motivated by the muon $g-2$ anomaly.
Taking rare $B$ and $K$ meson decay data into account,
we give upper limits on the $t\to cZ'$ branching ratio for the light $Z'$ case,
and discuss about its observability at the LHC.
We also scrutinize the possibility that the decay $K_L\to\pi^0Z'$ with $Z'\to \nu\bar\nu$
may lead to apparent violation of the usual Grossman--Nir bound of
${\cal B}(K_L\to\pi^0\nu\bar\nu)<1.4\times 10^{-9}$.
\begin{description}
\item[PACS numbers]
13.20.Eb 
13.20.He 
14.65.Ha 
14.70.Pw 
\end{description}
\end{abstract}

\maketitle


\section{\label{sec:Intro}Introduction\protect\\}

Measuring top quark properties is one of the major tasks at the Large Hadron Collider (LHC).
In particular, the large amount of $t\bar t$ events enables 
the search for exotic decay modes of the top quark.
In line with this, top quark decays via flavor-changing neutral currents (FCNCs) by emitting
the Standard Model (SM) gauge bosons or the Higgs boson have been under intense study
by the ATLAS and CMS experiments~\cite{Agashe:2014kda}.
Furthermore, top quark decays offer opportunities to directly search for New Physics,
e.g. a charged Higgs boson ($H^+$) in $t\to bH^+$ decay~\cite{Agashe:2014kda}.

For the down-sector counterpart, the decay properties of bottom quarks, 
i.e. $B$ mesons, have been investigated thoroughly in the last three decades.
So far, there are no significant deviations from SM in the vast amount of 
B factory and LHC data.
From LHC Run 1 data, however, the LHCb experiment has found two tantalizing hints 
for New Physics beyond SM (BSM) in $b\to s$ transitions.
%
One is the $\sim 3\sigma$ tension with SM in the angular analysis of
$B^0\to K^{*0}\mu^+\mu^-$ decay,
namely the $P_5'$ anomaly~\cite{Aaij:2013qta,LHCb:2015dla}.
The other is a violation of lepton flavor universality observed in
$B^+\to K^+\ell^+\ell^-$ ($\ell=e$ or $\mu$) decay rates, 
indicating a 2.6$\sigma$ discrepancy from the SM prediction,
i.e. the $R_K$ anomaly~\cite{Aaij:2014ora}.
Although these may be due to statistical fluctuations and/or 
underestimated hadronic uncertainties, model-independent studies for
possible BSM effects have revealed that both the 
$P_5'$~\cite{Descotes-Genon:2013wba,Altmannshofer:2013foa,Beaujean:2013soa,Horgan:2013pva,Hurth:2013ssa}
and $R_K$~\cite{Alonso:2014csa,Hiller:2014yaa,Ghosh:2014awa} anomalies
can be better explained by adding a new contribution to $C_9^\mu$,
the Wilson coefficient of the effective operator $(\bar s_L\gamma_\alpha b_L)(\bar \mu\gamma^\alpha \mu)$.
Surprisingly, the required amount of New Physics contribution to $C_9^\mu$
for the two anomalies is similar~\cite{Hurth:2014vma,Altmannshofer:2014rta}.

The extra contribution to $C_9^\mu$ can be generated by
a new massive gauge boson $Z'$ that couples to the muon vector current.
In Ref.~\cite{Altmannshofer:2014cfa}, 
a $Z'$ model was constructed based on the gauged $L_\mu-L_\tau$ symmetry~\cite{He:1990pn}, 
the difference between the muon and tauon numbers.
As an effective theory, the model permits $Z'$ couplings to the SM quark currents
through~\cite{Fox:2011qd} higher dimensional operators.
Ref.~\cite{Altmannshofer:2014cfa} gave a viable UV-complete model 
by introducing new vector-like quarks which mix with SM quarks.
The model predicts an effective $tcZ'$ coupling on the same footing, 
leading to $t\to cZ'$ decay if the $Z'$ is lighter than the top.
Such a light $Z'$ is allowed, as it is well hidden with no direct couplings to SM particles,
except the muon, tau and the associated neutrinos.
The $t\to cZ'$ decay, followed by $Z'\to \mu^+\mu^-/\tau^+\tau^-$,
opens up a new window to search for the $Z'$~\cite{Jung:2009jz}.

%
Another phenomenological motivation to introduce the gauged
$L_\mu-L_\tau$ symmetry comes from the long-standing $\sim 3\sigma$ discrepancy in
the muon anomalous magnetic moment $a_\mu\equiv(g_\mu-2)/2$ between
experimental data and SM predictions~\cite{Jegerlehner:2009ry}.
The radiative correction by the $Z'$ loop can provide~\cite{Baek:2001kca}
the right amount of shift, $\Delta a_\mu \sim 3\times 10^{-9}$, to match with data.
It was found~\cite{Altmannshofer:2014pba} recently, however, that 
the $Z'$ effect is severely constrained by the neutrino trident production 
$\nu_\mu N\to \nu_\mu N\mu^+\mu^-$ process:
the good agreement between experimental data and SM prediction excludes
a large portion of the parameter region that could explain the muon $g-2$ anomaly,
leaving only the rather {\it light} $Z'$ case, 
\begin{align}
m_{Z'}\lesssim 400~\MeV \quad ({\rm muon}~g-2). \label{eq:mZp-g-2}
\end{align}

Given this constraint, the muon $g-2$ anomaly and
the $b\to s$ transition anomalies cannot be explained simultaneously,
as the latter requires a {\it heavy} $Z'$ with mass suitably above $m_b$
to generate the contact $(\bar sb)(\bar\mu\mu)$ interaction for $C_9^\mu$.
But it is still interesting to investigate connections with the quark currents even for
the light $Z'$ case that satisfies Eq.~(\ref{eq:mZp-g-2}): the $t\to cZ'$ decay, 
followed by $Z'\to\mu^+\mu^-$, would exhibit the interesting collider signature of
collimated opposite-sign muons from a highly boosted $Z'$.
Moreover, the mass range of Eq.~(\ref{eq:mZp-g-2}) implies that the $Z'$ can be directly
produced in $B$ and $K$ meson decays.
We have pointed out in Ref.~\cite{Fuyuto:2014cya} that such a $Z'$ 
with mass around $m_{\pi^0}$ can evade $K^+\to\pi^+Z'$ searches, 
but cause $K_L\to\pi^0Z'\ (\to \nu\bar\nu)$ with rate exceeding
the commonly perceived Grossman--Nir (GN) bound~\cite{Grossman:1997sk}
of ${\cal B}(K_L\to\pi^0\nu\bar\nu)<1.4\times 10^{-9}$.
Note that a very light $Z'$ of $L_\mu-L_\tau$ 
might explain~\cite{Araki:2014ona} the PeV-scale cosmic neutrino spectrum
observed by IceCube~\cite{Aartsen:2014gkd}.

In this paper, we investigate how large the $t\to cZ'$ decay rate
can be,~\footnote{
Due to constraints from $D$ meson mixing and decay data,
$t\to cZ'$ and $t\to uZ'$ cannot be simultaneously large.
We concentrate on possibilities for large $t\to cZ'$ rates in this paper.
} 
based on the gauged $L_\mu -L_\tau$ model of Ref.~\cite{Altmannshofer:2014cfa}.
We consider the two well-motivated $Z'$ mass ranges:
(i) the heavy $Z'$ scenario with $m_b \lesssim m_{Z'} <m_t-m_c$, 
which is motivated by the $P_5'$ and $R_K$ anomalies;
(ii) the light $Z'$ scenario with $m_{Z'} \lesssim 400$ MeV, 
which is motivated by the muon $g-2$ anomaly.
The former was already sketched in Ref.~\cite{Altmannshofer:2014cfa}.
It was pointed out that the right-handed $tcZ'$ coupling is unconstrained
from $B$ and $K$ meson data and can lead to the $t\to cZ'$ decay
with $\sim 1$\% branching ratio.
We revisit their result by updating $b\to s$ transition data with a correction
to the $t\to cZ'$ formula.
On the other hand, Scenario (ii) was not studied in Ref.~\cite{Altmannshofer:2014cfa},
but clearly exhibits rather different phenomenology compared to Scenario (i).
As the on-shell $Z^\prime$ could be produced by $B$ and $K$ meson decays,
the meson decay rates could be hugely enhanced, and even the right-handed $tcZ'$
coupling is constrained by data at one-loop level.
Scenario (ii) is further divided into two categories:
(ii-a) $2m_\mu < m_{Z'}\lesssim 400$ MeV; (ii-b) $m_{Z'}< 2m_\mu$.
In Scenario (ii-b), the $Z'$ decays only into neutrinos, rendering
$t\to cZ'$ searches at the LHC difficult,
but it gives interesting implications for rare kaon decays as mentioned above.

This paper is organized as follows.
We recapitulate in Sec.~\ref{sec:heavy} the model of Ref.~\cite{Altmannshofer:2014cfa},
then study $t\to cZ'$ in Scenario (i) for heavy $Z'$ motivated by the $b\to s$ anomalies.
We then discuss the observability of $t\to cZ'$ decay at the LHC with $Z'\to \mu^+\mu^-$.
In Sec.~\ref{sec:light-mumu} and \ref{sec:light-nunu}, we study $t\to cZ'$ for light $Z'$
motivated by the muon $g-2$ anomaly.
We consider Scenario (ii-a) in Sec.~\ref{sec:light-mumu}, 
where the $Z'$ mass is above dimuon threshold.
We give formulas for FCNC $B$ and $K$ decays,
collect relevant rare $B$ and $K$ decay data, 
then give upper limits on $t\to cZ'$ branching ratio.
In Sec.~\ref{sec:light-nunu}, we consider Scenario (ii-b) 
where the $Z'$ is below the dimuon threshold.
After giving upper limits on $t\to cZ'$ branching ratios,
we discuss a special implication for rare kaon decay experiments,
expanding the discussion of Ref.~\cite{Fuyuto:2014cya}.
Sec.~\ref{sec:disc} is devoted to discussion and conclusions.
In Appendix~\ref{app:B2Kpimumu}, we give 
the decay distribution for $B^0\to K^{*0}Z'\to K\pi \mu^+\mu^-$ four-body decay 
to estimate the efficiency at LHCb.
In Appendix~\ref{app:loop}, loop functions used in our analysis are given.

\section{$P'_5$- and $R_K$-motivated $Z'$ \label{sec:heavy}}

\subsection{Model \label{subsec:model}}

We first recapitulate the model introduced in Ref.~\cite{Altmannshofer:2014cfa}.
A new Abelian gauge group U(1)$'$ is introduced that gauges the $L_\mu -L_\tau$ symmetry.
This U(1)$'$ symmetry is spontaneously broken by the vacuum expectation value (v.e.v.)
of a scalar field $\Phi$, which is charged under U(1)$'$ but singlet under the SM gauge group.
The mass of $Z'$ is given then by $m_{Z'}=g'v_\Phi$, where $g'$ is the U(1)$'$ gauge coupling
and $v_\Phi=\sqrt{2}\langle\Phi\rangle$ is the v.e.v. of $\Phi$.
We adopt the convention where the covariant derivative acting on $\Phi$ is given by
$D_\alpha=\partial_\alpha +ig'{\cal Q}'_\Phi Z'_\alpha$, with ${\cal Q}'_\Phi=+1$ the
U(1)$'$ charge of $\Phi$.
The $Z'$ couples to leptons via
\begin{align}
\mathcal{L}\supset 
-g'Z'_\alpha\left(\bar\mu\gamma^\alpha\mu +\bar\nu_{\mu L}
 \gamma^\alpha \nu_{\mu L} -\bar\tau\gamma^\alpha\tau -\bar\nu_{\tau L}
 \gamma^\alpha \nu_{\tau L}\right). \label{eq:gauge-coup}
\end{align}
We set the kinetic mixing between the U(1)$'$ and U(1)$_Y$ gauge bosons
to be zero throughout this paper.

In order to induce the effective $Z'$ couplings to the SM quark currents,
new vector-like quarks, which mix with the SM quarks, are introduced:
$Q_L=(U_L,D_L)$, $U_R$, $D_R$,
which replicates one generation of SM quarks,
and chiral partners $\tilde Q_R=(\tilde U_R, \tilde D_R)$, $\tilde U_L$, $\tilde D_L$.
Unlike the SM quarks, the new vector-like quarks are charged under U(1)$'$,
with charges ${\cal Q}'_Q=+1$ for $Q\equiv Q_L+\tilde Q_R$ and
${\cal Q}'_U={\cal Q}'_D=-1$ for $U\equiv \tilde U_L+U_R$ and $D\equiv \tilde D_L +D_R$.
The mass term for the vector-like quarks is given by
\begin{align}
-\mathcal L_{m}
=m_Q \bar Q Q +m_{U}\bar{U} U +m_{D}\bar{D} D,
\end{align}
where the three mass parameters are taken to be real without loss of generality.
The Yukawa mixing term between the vector-like quarks and SM quarks is given by
\begin{align}
-\mathcal L_{\rm mix}
&= \Phi \sum_{i=1}^3 \left( \bar{\tilde U}_R Y_{Qu_i}u_{iL}
  +\bar{\tilde D}_R Y_{Qd_i}d_{iL} \right) \notag\\
& +\Phi^\dagger\sum_{i=1}^3 \left( \bar{\tilde{U}}_L Y_{Uu_i}u_{iR}
 +\bar{\tilde{D}}_L Y_{Dd_i}d_{iR} \right)+{\rm h.c.}
\label{eq:Y-mix}
\end{align}
Here, SU(2)$_L$ symmetry imposes
\begin{align}
Y_{Qu_i}=\sum_{j=1}^3 V_{u_id_j}^* Y_{Qd_j}, \label{eq:Y-SU(2)}
\end{align}
for $i=1,2,3$, where $V_{u_id_j}$ is an element of the 
Cabibbo-Kobayashi-Maskawa (CKM) matrix.

\begin{figure}[t!]
{
 \includegraphics[width=80mm]{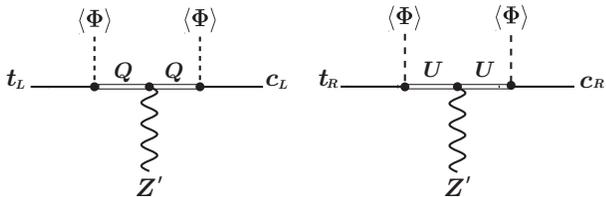}
}
\caption{Diagrams that induce effective $tcZ'$ couplings.
} \label{fig:tcZp}
\end{figure}

Integrating out the heavy vector-like quarks,
one obtains the effective $Z'$ couplings to the SM quarks as
\begin{align}
\mathcal L_{\rm eff}
&\supset 
 -Z'_\alpha \sum_{i,j=1}^3 \Big(g_{u_iu_j}^L \bar u_{iL}\gamma^\alpha u_{jL}
 +g_{u_iu_j}^R \bar u_{iR}\gamma^\alpha u_{jR} \notag\\
&\quad +g_{d_id_j}^L \bar d_{iL}\gamma^\alpha d_{jL}
 +g_{d_id_j}^R \bar d_{iR}\gamma^\alpha d_{jR} \Big), \label{eq:Leff-Z'}
\end{align}
where
\begin{align}
g_{u_iu_j}^L &= +g^\prime \frac{Y_{Qu_i}^*Y_{Qu_j}v_\Phi^2}{2m_Q^2},~
 g_{u_iu_j}^R = -g^\prime \frac{Y_{Uu_i}^*Y_{Uu_j}v_\Phi^2}{2m_U^2}, \notag\\
g_{d_id_j}^L &= +g^\prime \frac{Y_{Qd_i}^*Y_{Qd_j}v_\Phi^2}{2m_Q^2},~
 g_{d_id_j}^R = -g^\prime \frac{Y_{Dd_i}^*Y_{Dd_j}v_\Phi^2}{2m_D^2}. \label{eq:eff-coup}
\end{align}
The effective couplings to $t\to c$ currents, for instance,  are
generated by the diagrams shown in Fig.~\ref{fig:tcZp}.

After integrating out $Q,U,D$, the Yukawa mixing couplings in Eq.~(\ref{eq:Y-mix})
induce also the effective couplings of $\Phi$ to the SM quark bilinears.
In particular, the physical mode of the $\Phi$ field, $\phi$, 
may couple to the top and charm quarks, 
hence the top FCNC decay $t\to c\phi$ can occur, 
if the $\phi$ boson is lighter than the top.
In this paper, we concentrate on the decay $t\to cZ'$, assuming that the $\phi$
is heavier than the top.
Phenomenology of the U(1)$'$ Higgs boson $\phi$ will be 
investigated elsewhere~\cite{HK:phi}.

\subsection{$P_5'$ and $R_K$ anomalies}

The effective $bsZ'$ couplings in Eq.~(\ref{eq:eff-coup}), in combination with
Eq. (\ref{eq:gauge-coup}), induce extra contributions to $b\to s\mu^+\mu^-$ decays.
Assuming the $Z'$ is heavy compared to the $B$ meson mass scale,
one may integrate out the $Z'$ and obtain BSM contributions to
the $b\to s$ effective Hamiltonian
\begin{align}
 \Delta C_9^{\mu}(\bar s\gamma_\alpha P_L b)(\bar\mu\gamma^\alpha\mu)
 +C_9^{\prime\mu}(\bar s\gamma_\alpha P_R b)(\bar\mu\gamma^\alpha\mu),
\end{align}
with the Wilson coefficients
\begin{align}
\Delta C_9^{\mu}&\simeq +\frac{Y_{Qs}^*Y_{Qb}}{2m_Q^2}, \quad
C_9^{\prime\mu} \simeq -\frac{Y_{Ds}^*Y_{Db}}{2m_D^2}, \label{eq:WC}
\end{align}
where $g'$ and $v_\Phi$ cancel out in the final expressions.
Note that the electron counterparts are unchanged, i.e. $\Delta C_9^{e}=C_9^{\prime e}=0$,
resulting in the violation of lepton flavor universality in $b\to s\ell^+\ell^-$ 
between $\ell=\mu,e$.

In Ref.~\cite{Altmannshofer:2014cfa}, the values $\Delta C_9^{\mu}\simeq
-(35~{\rm TeV})^{-2}$, $C_9^{\prime\mu} \simeq +(35~{\rm TeV})^{-2}$
from a global analysis~\cite{Altmannshofer:2013foa} of $b\to s$ data was adopted
as a solution for the $P_5'$ anomaly~\cite{Aaij:2013qta}.
Since then, the $R_K$ anomaly~\cite{Aaij:2014ora} has emerged, while $P_5'$
has been updated with the 3 fb$^{-1}$ dataset by LHCb~\cite{LHCb:2015dla}.
A recent global analysis~\cite{Altmannshofer:2014rta},
after the $P_5'$ update, found the best fit values $\Delta C_9^{\mu}
\simeq -(34~{\rm TeV})^{-2}$, $C_9^{\prime\mu} \simeq +(54~{\rm TeV})^{-2}$.
We remark that this fit does not include $b\to se^+e^-$ modes,
while $R_K$ is defined by the ratio
$R_K\equiv {\cal B}(B^+\to K^+\mu^+\mu^-)/{\cal B}(B^+\to K^+e^+e^-)$.
The measured $R_K$ value~\cite{Aaij:2014ora} is better explained by
$C_9^{\prime\mu}\sim 0$ with a $\Delta C_9^{\mu}$ value
similar to the above best fit value, which is still within the 1$\sigma$ ellipse
of the allowed region in Ref.~\cite{Altmannshofer:2014rta}.
For illustration, therefore, we take the following values as a reference point
to solve the $b\to s$ anomalies:
\begin{align}
\Delta C_9^{\mu}\simeq -(34~{\rm TeV})^{-2}, \quad
C_9^{\prime\mu} \sim 0. \label{eq:b2s-sol-1}
\end{align}

Interpreted within the $Z'$ model, the values of Eq.~(\ref{eq:b2s-sol-1}) 
correspond to~\footnote{
Note that the sign for $Y_{Qb}Y_{Qs}^*$ is opposite the one 
in Ref.~\cite{Altmannshofer:2014cfa}.
}
\begin{align}
m_Q &\simeq 24~{\rm TeV}\times (-Y_{Qb}Y_{Qs}^*)^{1/2}, \label{eq:b2s-sol-2}
\end{align}
with real and negative $Y_{Qb}Y_{Qs}^*$, and $Y_{Db}Y_{Ds}^*/m_D^2\sim 0$.
The latter implies the vector-like quark $D$ is decoupled.
For hierarchical Yukawa couplings, $Y_{Qb}=1$, $Y_{Qs}=-\lambda^2$
with $\lambda\simeq 0.23$, Eq.~(\ref{eq:b2s-sol-2}) implies $m_Q\simeq 5.5$ TeV. 

%
\subsection{Constraints on $v_\Phi$ \label{sec:vphi-bound}}

Before entering the discussion on the $t\to cZ'$ decay, we summarize constraints on
$v_\Phi$, the v.e.v. of the U(1)$'$ Higgs field, which is of great importance to $t\to cZ'$.

The most significant constraint comes from neutrino trident production, i.e.
$\nu_\mu N\to \nu_\mu N\mu^+\mu^-$,
with the normalized cross section~\cite{Altmannshofer:2014cfa}
\begin{align}
\frac{\sigma}{\sigma_{\rm SM}}
\simeq \frac{1+(1+4s_W^2 +2v^2/v_\Phi^2)^2}{ 1+(1+4s_W^2 )^2 },
\end{align}
for heavy $Z'$.
Using CCFR data \cite{Mishra:1991bv}, which imply
$\sigma_{\rm exp}/\sigma_{\rm SM}=0.82\pm 0.28$,
we obtain the $2\sigma$ lower bound
\begin{align}
v_\Phi \gtrsim 540~{\rm GeV}~{\rm for}~m_{Z'}\gtrsim 10~\GeV.
\label{eq:NTP-heavy}
\end{align}
For a fixed $m_{Z'}$, this constraint can be translated into an upper bound on
$g'=m_{Z'}/v_\Phi$, i.e., $g' \lesssim 0.09 \times (m_{Z'}/50~{\rm GeV})$.
For $m_{Z'}\lesssim 10$ GeV, the constraint is softened \cite{Altmannshofer:2014pba},
as we will see in the next section.

For $Z'$ lighter than the $Z$ boson, the coupling $g'$ is also constrained
by $Z\to 4\ell$ searches at the LHC.
An analysis~\cite{Altmannshofer:2014cfa,Altmannshofer:2014pba}
that utilizes
the Run 1 result of ATLAS~\cite{Aad:2014wra} 
found that the $Z\to 4\ell$~\cite{Harigaya:2013twa} provides slightly tighter constraints
than the neutrino trident production for $9\,\GeV \lesssim m_{Z'}\lesssim 50\,\GeV$.
The strongest bound $g' \lesssim 0.01$ is attained for $m_{Z'}\sim 10$ GeV,
leading to $v_\Phi \gtrsim 1000$ GeV around this mass value.
This loses sensitivity for lower $Z'$ masses due to the cut applied in
the experimental analysis.

For $m_{Z'}\gg m_\mu$, the $Z'$ contribution to muon $g-2$ is given by
$\Delta a_\mu\simeq m_\mu^2/(12\pi^2v_\Phi^2)$~\cite{Baek:2001kca}.
To explain the discrepancy between experiment and theory,
$\Delta a_\mu = (2.9\pm 0.9)\times 10^{-9}$~\cite{Jegerlehner:2009ry},
one needs $160~\GeV \lesssim v_\Phi \lesssim 220$ GeV.
This range is excluded by the constraint from the neutrino trident production,
Eq.~(\ref{eq:NTP-heavy}).
The case for the light $Z'$ will be discussed in the next section.

The effective $bsZ'$ coupling induces $B_s$ mixing, which provides
an upper bound on $v_\Phi$.
The modification to the $B_s$ mixing amplitude is given by~\cite{Altmannshofer:2014cfa}
\begin{align}
&\frac{M_{12}}{M_{12}^{\rm SM}} \simeq 1 +(Y_{Qb}Y_{Qs}^*)^2
 \left(\frac{v_\Phi^2}{m_Q^4} +\frac{1}{16\pi^2}\frac{1}{m_Q^2}\right) \notag\\
&\qquad\qquad \times \left[ \frac{g_2^4}{16\pi^2}\frac{1}{m_W^2}(V_{ts}^*V_{tb})^2S_0
  \right]^{-1}, \label{eq:B-mixing-heavy}
\end{align}
where $S_0\simeq 2.3$ and the $D$ quark effects are decoupled, given the $b\to s$
transition data.
It is useful to eliminate~\cite{Crivellin:2015mga} the dependence on $Y_{Qb}Y_{Qs}^*/m_Q^2$
in terms of $\Delta C_9^\mu$ of Eq.~(\ref{eq:WC}).
Then, allowing BSM effects up to 15\%~\cite{Altmannshofer:2014cfa},
we find the upper bound
\begin{align}
v_\Phi\lesssim 5.6~\TeV \left( \frac{(34~\TeV)^{-2}}{|\Delta C_9^\mu|} \right).
 \label{eq:B-mixing-vphi}
\end{align}
We have neglected the $1/m_Q^2$ term in Eq.~(\ref{eq:B-mixing-heavy}),
which is numerically valid for $m_Q\lesssim 10$ TeV.
For larger $m_Q$, the bound gets gradually stronger, e.g.
$v_\Phi\lesssim 5.4\ (3.9)$ TeV for $m_Q=20\ (50)$ TeV, with $\Delta C_9^\mu$
satisfying Eq.~(\ref{eq:b2s-sol-1}).


%
The constraint from kaon mixing can be avoided by assuming the mixing of
$Q$ and $D$ quarks with $d$ quark is suppressed: $Y_{Qd}\simeq Y_{Dd}\simeq 0$.
Although this assumption leads to $Y_{Qu}\simeq \lambda Y_{Qs}$
via Eq.~(\ref{eq:Y-SU(2)}), hence a new contribution to $D$ mixing, 
$B_s$ mixing still gives the strongest constraint~\cite{Altmannshofer:2014cfa}.
We further set $Y_{Uu}\simeq 0$ to switch off the right-handed $c\to u$ current contribution
to $D$ mixing, in order to pursue the possibility of large $t\to cZ'$ rate.

\subsection{Branching ratio for $t\to cZ'$}

We now turn to $t\to cZ^\prime$ decay.
With the effective $tcZ'$ couplings in Eq.~(\ref{eq:eff-coup}),
the decay rate is given by
\begin{align}
&\Gamma(t\to cZ^\prime) \notag\\
&=\frac{m_t}{32\pi}\lambda^{1/2}(1,x_c,x^\prime)
 \bigg[ \left( |g_{ct}^L|^2+|g_{ct}^R|^2 \right) \big[ 1+x_c-2x^\prime \notag\\
 &\quad + {(1-x_c)^2}/{x^\prime} \big]
  -12{\rm Re}(g_{ct}^Rg_{ct}^{L*})\sqrt{x_c} \bigg],
\end{align}
where $x_c\equiv m_c^2/m_t^2$, $x' \equiv m_{Z'}^2/m_t^2$ and
\begin{align}
\lambda(x,y,z) \equiv x^2+y^2+z^2-2(xy+yz+zx). \label{eq:lambda-func}
\end{align}
Taking the ratio with the $t\to bW$ rate, the $t\to cZ^\prime$ branching ratio is given by
\begin{align}
\mathcal B(t\to cZ^\prime)
&\simeq \frac{(1-x^\prime)^2(1+2x^\prime)}{2(1-x_W)^2(1+2x_W)} \notag\\
&\times\left( |Y_{Qt} Y_{Qc}^*|^2\frac{v^2v_\Phi^2}{4m_Q^4}
+|Y_{Ut} Y_{Uc}^*|^2\frac{v^2v_\Phi^2}{4m_{U}^4} \right), \label{eq:t2cZp}
\end{align}
where $x_W\equiv m_W^2/m_t^2$,
and we have set $m_c^2/m_t^2$, $m_b^2/m_t^2 \to 0$.
Note that our result is a factor of four smaller than the one shown in
Ref. \cite{Altmannshofer:2014cfa}.

The first term in Eq.~(\ref{eq:t2cZp}) is induced by the left-handed $t\to c$ current
(first diagram in Fig.~\ref{fig:tcZp}),
which is related to the left-handed $b\to s$ current by SU(2)$_L$ symmetry.
Neglecting Cabibbo suppressed terms, Eq.~(\ref{eq:Y-SU(2)}) implies
$Y_{Qt}\sim Y_{Qb}$, $Y_{Qc}\sim Y_{Qs}$.
One may eliminate the $Y_{Qt}Y_{Qc}^*/m_Q^2$ dependence in Eq.~(\ref{eq:t2cZp})
by using $\Delta C_9^\mu$ [Eq.~(\ref{eq:WC})]
to rewrite the left-handed current contribution 
\begin{align}
{\cal B}(t\to cZ')_{\rm LH}
&\simeq \frac{(1-x^\prime)^2(1+2x^\prime)}{2(1-x_W)^2(1+2x_W)}
 |\Delta C_9^\mu|^2v^2v_\Phi^2 \label{eq:t2cZp-LH-1}.
\end{align}
Applying the lower bound, Eq.~(\ref{eq:NTP-heavy}), 
and upper bound, Eq.~(\ref{eq:B-mixing-vphi}), on $v_\Phi$,
we obtain the allowed range for left-handed current contribution:
\begin{align}
0.7\times 10^{-8} \lesssim {\cal B}(t\to cZ')_{\rm LH}
 \lesssim 0.8\times 10^{-6},
  \label{eq:t2cZp-LH-2}
\end{align}
for a $Z'$ mass sufficiently below the kinematic threshold.
Note that the lower limit assumes the central value of $\Delta C_9^\mu$ from
the global fit in Eq.~(\ref{eq:b2s-sol-1}),
while the upper limit is insensitive to $\Delta C_9^\mu$ as the dependence cancels out.
The branching ratio can be slightly larger than the upper value quoted
in Ref.~\cite{Altmannshofer:2014cfa}, i.e. few $\times 10^{-7}$.
This is because the $B_s$-mixing constraint on $v_\Phi$, Eq. (\ref{eq:B-mixing-vphi}),
is weakened due to the decoupling of $D$ quark effects, which is favored by
the measured $R_K$ value.

On the other hand, the second term in Eq.~(\ref{eq:t2cZp}), induced by
the right-handed $t\to c$ current (second diagram in Fig.~\ref{fig:tcZp}),
is not related to FCNCs in the down-type quark sector.
Treating $m_U$, $Y_{Ut}$ and $Y_{Uc}$ as free parameters, the right-handed current
contribution can be easily enhanced over the left-handed current contribution.
To see how large it can be, we introduce a mixing parameter between the vector-like
quark $U$ and $t_R$ or $c_R$ as
\begin{align}
\delta_{Uq} \equiv \frac{ Y_{Uq} v_\Phi }{\sqrt{2}m_U}, \quad\quad (q=t,c)
\end{align}
and recast the right-handed current contribution as
\begin{align}
\mathcal B(t\to cZ^\prime)_{\rm RH}
&\simeq \frac{(1-x^\prime)^2(1+2x^\prime)}{2(1-x_W)^2(1+2x_W)}\frac{v^2}{v_\Phi^2}
 \left| \delta_{Ut}\delta_{Uc}^* \right|^2. \label{eq:t2cZ'-RH}
\end{align}
For fixed values of $\delta_{Ut}$ and $\delta_{Uc}$,
this can be enhanced by lowering the value of $v_\Phi$, which is
bounded from below by neutrino trident production [Eq.~(\ref{eq:NTP-heavy})].
Taking reasonably large mixing parameters $\delta_{Ut}=\delta_{Uc} \simeq \lambda$
for illustration, Eq.~(\ref{eq:NTP-heavy}) implies
\begin{align}
\mathcal B(t\to cZ^\prime)_{\rm RH}\lesssim 3\times 10^{-4} \quad
(\delta_{Ut}=\delta_{Uc} \simeq \lambda).
\end{align}
This is smaller than the value in Ref.~\cite{Altmannshofer:2014cfa},
i.e. $\sim 1$\%, partially because of the correction factor $1/4$ in Eq.~(\ref{eq:t2cZp}).
In the corrected formula, the $\sim 1$\% branching ratio requires
rather large mixing parameters: $\delta_{Ut}\sim \delta_{Uc} \sim 0.5$.

\begin{figure}[t!]
{
 \includegraphics[width=75mm]{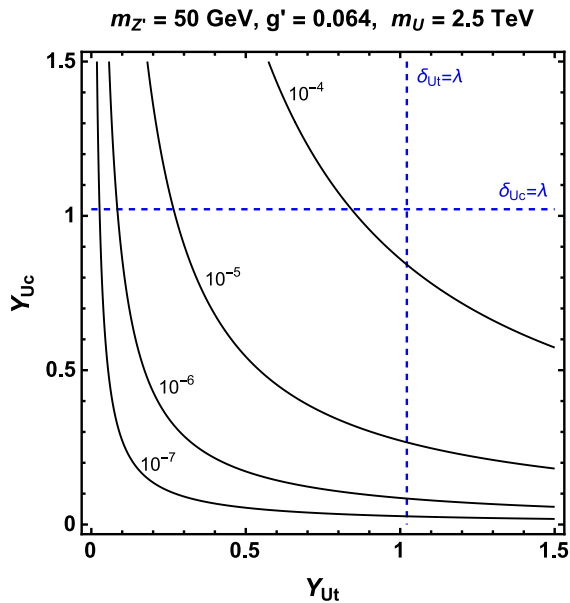}
}
\caption{
Contours for ${\cal B}(t\to cZ')_{\rm RH}$ in the $(Y_{Ut},\;Y_{Uc})$ plane
for $m_{Z'}=50$ GeV, $g'=0.064$, $v_\Phi = 780$ GeV, and $m_U=2.5$ TeV.
In estimating ${\cal B}(t\to cZ')$, only the contribution from
second term in Eq.~(\ref{eq:t2cZp}), i.e. right-handed $t\to c$ current, is included.
The vertical (horizontal) blue dashed line marks the value of $Y_{Ut}$ ($Y_{Uc}$)
above which the mixing parameter $\delta_{Ut} (\delta_{Uc})$ exceeds $\lambda\simeq 0.23$.
} \label{fig:50_Y}
\end{figure}

In Fig.~\ref{fig:50_Y}, contours of ${\cal B}(t\to cZ')_{\rm RH}$ are
given in the $(Y_{Ut},\,Y_{Uc})$ plane
for $m_{Z'}=50$ GeV, $g'=0.064$, $v_\Phi = 780$ GeV, and $m_U=2.5$ TeV.
The vertical (horizontal) dashed lines mark the value of $Y_{Ut}$ ($Y_{Uc}$)
above which the mixing parameter $\delta_{Ut}\ (\delta_{Uc})$ exceeds $\lambda$.
These lines are placed as a rough indication for reasonable range of
Yukawa mixing couplings $Y_{Uq}$.
The figure illustrates that ${\cal B}(t\to cZ')_{\rm RH}$ can exceed $10^{-4}$ for
$Y_{Ut},Y_{Uc}\sim 1$, with $\delta_{Ut},\; \delta_{Uc}<\lambda$ satisfied.

%

Before turning to search at LHC, we give the partial widths for 
$Z'$ decay
\begin{align}
&\Gamma(Z'\to \ell^+\ell^-)
=\frac{g'^2}{12\pi}m_{Z'}\left[ 1-\frac{4m_\ell^2}{m_{Z'}^2}\right]^{\frac{1}{2}}
 \left[1+\frac{2m_\ell^2}{m_{Z'}^2} \right], \notag\\
&\Gamma(Z'\to \nu_\ell\bar\nu_\ell)=\frac{g'^2}{24\pi}m_{Z'}, \label{eq:Zp-width}
\end{align}
where $\ell=\mu,\;\tau$.
The approximate branching ratios are
\begin{align}
&{\cal B}_{\tau\tau}\simeq {\cal B}_{\mu\mu}\simeq
 {\cal B}_{\nu\nu}\simeq \frac{1}{3} \quad (2m_\tau\ll m_{Z^\prime}), \notag\\
&{\cal B}_{\mu\mu}\simeq {\cal B}_{\nu\nu}\simeq \frac{1}{2}
 \quad ( 2m_\mu \ll m_{Z^\prime}< 2m_\tau), \notag\\
&\mathcal B_{\nu\nu}= 1 \quad (m_{Z^\prime}<2m_\mu).
 \label{eq:Zp-BR}
\end{align}

\subsection{$t\to cZ'$ search at LHC}

The decay $t\to cZ'$ followed by $Z'\to\ell^+\ell^-$ ($\ell=\mu,\tau$)
can be searched for in $t\bar t$ events at the LHC.
It is similar to $t\to qZ$ ($q=u,c$) decay, which has been searched for 
by the ATLAS~\cite{Aad:2015uza} and CMS~\cite{Chatrchyan:2013nwa} experiments
using $t\bar t \to Zq+Wb$ with leptonically decaying $Z$ and $W$, resulting
in a final state with three charged leptons.
The $t\to cZ'$ decay with heavy $Z'$ should be searched for
in an analogous way, by modifying event selection criteria for 
an opposite sign charged lepton pair.

The current best limit on the $t\to qZ$ rate comes from the CMS analysis
with the full Run 1 dataset~\cite{Chatrchyan:2013nwa}, finding
${\cal B} (t\to qZ)< 5\times 10^{-4}$ at 95\% C.L.,
while the ATLAS~\cite{Aad:2015uza}, based on the 20.3 fb$^{-1}$ dataset of
the 8 TeV run, found ${\cal B}(t\to qZ)< 7\times 10^{-4}$ at 95\% C.L.
These limits should be improved with more data during the 13/14 TeV run of the LHC.
The expected limits at the 14 TeV LHC with 300 fb$^{-1}$ (3000 fb$^{-1}$) data are
${\cal B}(t\to qZ) < 2.7\times 10^{-4}~(1.0\times 10^{-4})$ for
CMS~\cite{CMS:2013zfa},~\footnote{
In the Snowmass White Paper~\cite{CMS:2013xfa}, a more optimistic value
$\sim 10^{-5}$ is quoted as the CMS sensitivity for $t\to qZ$
with 300 fb$^{-1}$ data at the 14 TeV LHC.
This was based on extrapolating from the 7 TeV result~\cite{Chatrchyan:2012hqa}.
The projections in Ref. \cite{CMS:2013zfa}, on the other hand, are based on
Monte Carlo analysis.
}
and ${\cal B}(t\to qZ) < 2.2\times 10^{-4}$ ($7\times 10^{-5}$) 
for ATLAS~\cite{ATLAS:2013hta,Agashe:2013hma}.

For illustration, we attempt a reinterpretation of the CMS limits for $t\to cZ$
to the case for $t\to cZ'$ by a simple scaling of $Z$ and $Z'$ decay branching
ratios into the charged leptons ($\ell=e,\mu$).
An advantage of the $t\to cZ'$ search is the larger $Z'$ branching ratio,
e.g., ${\cal B}(Z'\to \mu^+\mu^-)\simeq 1/3$ for $m_{Z'}\gg 2m_\tau$,
compared with ${\cal B}(Z\to \ell^+\ell^-)\simeq 0.07$ (summed over $e$ and $\mu$).
Multiplying the factor of ${\cal B}(Z\to \ell^+\ell^-)/{\cal B}(Z'\to \ell^+\ell^-)
\simeq 0.2$ to the current \cite{Chatrchyan:2013nwa} and future \cite{CMS:2013zfa}
CMS limits for $t\to cZ$, we infer sensitivities for $t\to cZ'(\to \mu^+\mu^-)$ by CMS as
\begin{align}
{\cal B} (t\to cZ')\lesssim \begin{cases}
 10^{-4} & {\rm (``CMS"~Run~1)}, \\
 5\times 10^{-5} & {\rm (``CMS"~300~fb^{-1})}, \\
 2\times 10^{-5} & {\rm (``CMS"~3000~fb^{-1})}
\end{cases} \label{eq:t2cZp-CMS}
\end{align}
for the heavy $Z'$ with $m_{Z'}\sim {\cal O}(10)$ GeV.
The scaling of the ATLAS limits gives similar results.
Therefore, the right-handed current induced $t\to cZ'$ might 
already be probed by Run 1 data (see Fig.~\ref{fig:50_Y}),
while the left-handed current contribution
seems to be beyond the sensitivity of LHC, 
even with 3000 fb$^{-1}$ data [see Eq.~(\ref{eq:t2cZp-LH-2})].

For light $Z'$ with $2m_\mu < m_{Z'}\lesssim 400$ MeV,
the scaling factor is slightly reduced as
${\cal B}(Z\to \ell^+\ell^-)/{\cal B}(Z'\to \ell^+\ell^-)\simeq 0.14$
due to larger $Z'\to \mu^+\mu^-$ branching ratio ($\simeq 1/2$).
For such a light $Z'$, however, the search strategy needs to be changed.
In particular, muon pairs produced by boosted $Z'$ bosons would be highly collimated,
while the existing $t\to qZ$ search requires events with isolated charged leptons.
Nevertheless, we adopt Eq.~(\ref{eq:t2cZp-CMS}) for the light $Z'$ case
as target values in the following analysis.


For the light $Z'$ with $m_{Z'}< 2m_\mu$, the $Z'$ decays only into
neutrino pairs. Thus, the search at the LHC would be quite challenging.
\footnote{
 The $t\to q$ ($q=u,c$) decay with missing energy has been discussed based on
dark matter models~\cite{Jia:2015uea,D'Hondt:2015jbs}. 
}

\section{Muon $g-2$ and $Z'$
 \label{sec:light-mumu}}

\begin{figure}[t!]
{
 \includegraphics[width=75mm]{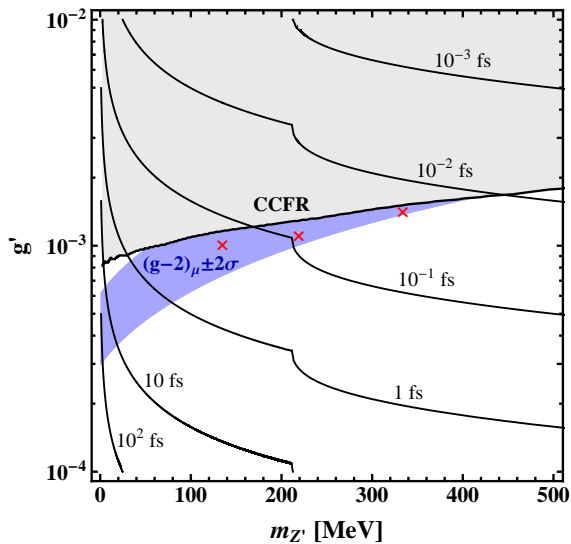}
}
\caption{
Lifetime of light $Z'$ with relevant constraints in the $(m_{Z'}, g')$ plane:
solid lines are labeled contours for $\tau_{Z'}$,
the blue-shaded band is the region favored by muon $g-2$ 
within $2\sigma$~\cite{Jegerlehner:2009ry},
the gray-shaded region is excluded by neutrino trident production~\cite{Mishra:1991bv}
and taken from Ref.~\cite{Altmannshofer:2014pba}.
The red crosses at $(m_{Z'}, g')= (135~\MeV, 10^{-3}), (219~\MeV, 1.1\times 10^{-3}), 
(334~\MeV, 1.4\times 10^{-3})$
indicate benchmark points adopted in our numerical study, as explained in text.
} \label{fig:lifetime}
\end{figure}

In this and following sections, 
we consider the {\it light} $Z'$ scenarios motivated by muon $g-2$ anomaly.
In Fig.~\ref{fig:lifetime}, we give the parameter region (the blue band) 
in the $(m_{Z'},g')$ plane that accounts~\cite{Baek:2001kca} for the discrepancy, 
$\Delta a_\mu = (2.9\pm 0.9)\times 10^{-9}$~\cite{Jegerlehner:2009ry},
taking $2\sigma$ error range.
The parameter space is strongly constrained~\cite{Altmannshofer:2014pba} by 
neutrino trident production, the gray-shaded exclusion region.
Thus, the $Z'$ boson of $L_\mu -L_\tau$ symmetry can explain
the muon $g-2$ anomaly only if $m_{Z'}\lesssim 400$ MeV, 
as given in Eq.~(\ref{eq:mZp-g-2}).
In this section, we consider the $t\to cZ'$ decay in the scenario of
\begin{align}
2m_\mu < m_{Z'}\lesssim 400 \MeV, \quad\quad [{\rm Scenario~(ii\mathchar`-a)}] \label{eq:s-ii-a}
\end{align}
which permits the $Z'\to \mu^+\mu^-$ decay.

%

The $Z'$ lifetime, $\tau_{Z'}$, estimated by summing up Eqs.~(\ref{eq:Zp-width}), 
is also given in Fig.~\ref{fig:lifetime} as the black-solid contours.
We see that $\tau_{Z'}\lesssim 0.1$ fs for the muon $g-2$ favored region above
the dimuon threshold. The decay length of a light $Z'$ with energy $E_{Z'}$ is given by
\begin{align}
\gamma c\tau_{Z'} &\simeq 0.4\,\mu {\rm m}
 \left[\frac{2}{N_{\rm eff}}\right]\left[\frac{10^{-3}}{g'}\right]^2
 \left[\frac{0.3\,\GeV}{m_{Z'}}\right]^2 \left[\frac{E_{Z'}}{10\,\GeV}\right], 
 \label{eq:decay-length}
\end{align}
where $N_{\rm eff}\simeq 2$ for $2m_\mu \ll m_{Z'}< 2m_\tau$,
and $N_{\rm eff}\simeq 1$ for $m_{Z'}< 2m_\mu$.
Thus, for $m_{Z'}\gtrsim 2m_\mu$, the $Z'$ that is motivated by muon $(g-2)$ 
decays promptly after production at colliders such as LHC and $B$ factories,
with branching ratios approximately shown in Eq.~(\ref{eq:Zp-BR}).
For $m_{Z'}<2m_\mu$, the lifetime can be significantly longer for extremely light ${Z'}$,
but its existence is simply felt as a missing energy (with no missing mass) 
in collider experiments, regardless of its decay.

\subsection{FCNC $B$ decays}

As the $Z'$ mass range in Eq.~(\ref{eq:s-ii-a}) is
too low to explain the $P_5'$ and $R_K$ anomalies,
we treat rare $B$ meson decay data as providing constraints on the effective $bsZ'$ coupling.
By SU(2)$_L$ symmetry, this also constrains the left-handed $tcZ'$ coupling.
In the light $Z'$ scenario, rare $B$ meson decays provide rather strong constraints,
and even the right-handed $tcZ'$ coupling becomes significantly constrained at one-loop level.
We also discuss rare kaon decay constraints on the latter.

\subsubsection{$B\to K^{(*)} Z'$ formulas}

The light $Z'$ can be produced directly in $B \to K^{(*)}Z'$ decays,
with $Z'\to \mu^+\mu^-/\nu\bar\nu$.
For the $bsZ'$ couplings of Eq. (\ref{eq:Leff-Z'}), the branching
ratio is given by\footnote{
We imply both $B^0\to K^0 Z'$ and $B^\pm \to K^\pm Z'$.
A similar convention applies to $B\to K^* Z'$. 
}
\begin{align}
&{\cal B}(\bar B\to \bar KZ') = \frac{\left|g_{sb}^L+g_{sb}^R\right|^2}{64\pi}
 \frac{m_B^3\beta^{3}_{BKZ'}}{m_{Z'}^2\Gamma_B}\left[f_+^{BK}(m_{Z'}^2)\right]^2, \label{eq:B2KZp}
\end{align}
where $f_+^{BK}$ is the $B\to K$ form factor and
\begin{align}
\beta_{XYZ}\equiv \lambda^{1/2}(1, m_Y^2/m_X^2, m_Z^2/m_X^2), \label{eq:beta-func}
\end{align}
with $\lambda(x,y,z)$ defined in Eq.~(\ref{eq:lambda-func}).
For $B\to K^*Z'$, the branching ratio can be expressed as~\cite{Oh:2009fm}
\begin{align}
{\cal B}(\bar B\to K^*Z') &=\frac{ \beta_{BK^*Z'}}{16\pi m_B\Gamma_B}
 \left( |H_0|^2 +|H_{+}|^2 +|H_{-}|^2 \right), \label{eq:B2KsZp}
\end{align}
where the helicity amplitudes $H_{0,\pm}$ are given by
\begin{align}
H_0 &= (g_{sb}^L-g_{sb}^R)\bigg[ -\frac{1}{2}(m_B+m_{K^*})\xi A_1(m_{Z'}^2) \notag\\
 &\quad +\frac{m_{K^*}m_{Z'}}{m_B+m_{K^*}}(\xi^2-1)A_2(m_{Z'}^2) \bigg], \notag\\
H_{\pm} &= \frac{1}{2}(g_{sb}^L-g_{sb}^R) (m_B+m_{K^*})A_1(m_{Z'}^2)  \notag\\
 &\quad \pm (g_{sb}^L+g_{sb}^R) \frac{m_{K^*}m_{Z'}}{m_B+m_{K^*}}\sqrt{\xi^2-1}
 V(m_{Z'}^2), \label{eq:Hm}
\end{align}
with $\xi \equiv (m_B^2-m_{K^*}^2-m_{Z'}^2)/(2m_{K^*}m_{Z'})$, and
$A_1,A_2, V$ are $B\to K^*$ form factors.
For the form factor numerical values, we adopt the fit formulas
from light-cone sum rule calculations~\cite{Ball:2004ye,Ball:2004rg}.
As the $Z'$ couples to the muon through the vector current,
there is no new physics contribution to $B_s\to \mu^+\mu^-$.

For later convenience, we provide numerical expressions of the $B^+\to K^+Z'$ and
$B^0\to K^{*0}Z'$ branching ratios for $m_{Z'}\lesssim 400$ MeV:
\begin{align}
\mathcal B(B^+\to K^+ Z')
&\simeq 2.2\times 10^{12}|g_{sb}^L+g_{sb}^R|^2 \left(\frac{300~{\MeV}}{m_{Z'}}\right)^2,\notag\\
\mathcal B(B^0\to K^{*0} Z')
&\simeq 2.4\times 10^{12} |g_{sb}^L-g_{sb}^R|^2 \left( \frac{300~{\MeV}}{m_{Z'}}\right)^2
\notag\\ &\quad +1.4\times 10^{10} |g_{sb}^L+g_{sb}^R|^2. 
\label{eq:B2KZp-num}
\end{align}
Note that the $(m_B/m_{Z'})^{2}$ enhancement comes from the longitudinally polarized $Z'$.

\subsubsection{$B\to K^{(*)}\mu^+\mu^-$ data}

With $\sim 50$\% $Z'$ decaying into muon pairs, 
$B\to K^{(*)}Z'$ decays would leave footprints in the dimuon mass
($q^2=m_{\mu\mu}^2$) spectra of $B\to K^{(*)}\mu^+\mu^-$ decays.
As the SM prediction~\cite{Beneke:2001at} is not reliable for $q^2 < 1$ GeV$^2$, 
one challenge for low-mass new boson search is the estimation of SM background.
Instead, one could take a data-based approach~\cite{Williams:2015xfa} 
by searching for a narrow peak in the dimuon spectrum.
With full 3.0 fb$^{-1}$ Run 1 data, the LHCb experiment has 
performed~\cite{Aaij:2015tna} such a dedicated search 
for a new hidden-sector boson $\chi$ in $B^0\to K^{*0}\chi$ with $\chi\to \mu^+\mu^-$.
Scanning the dimuon spectrum for 214 MeV $\leq m_{\mu\mu} \leq$ 4350 MeV
and finding no evidence for a signal, upper limits of ${\cal O}(10^{-9})$ on
${\cal B}(B^0\to K^{*0}\chi){\cal B}(\chi\to \mu^+\mu^-)$ are set 
for most of the $m_\chi$ range with $\tau_\chi \leq 100$ ps.

As the LHCb analysis \cite{Aaij:2015tna} assumed $\chi$ to be scalar,
the upper limits could be different for the $Z'$ case due to difference in efficiency.
We estimated the ratio of efficiency for vector vs scalar boson, 
based on information in supplemental material of Ref.~\cite{Aaij:2015tna}, 
and confirmed that the change is rather small, 
varying within 0--20\% in the mass range of our interest.
(See Appendix~\ref{app:B2Kpimumu} for detail.)
Hence, we can apply directly the limits in Ref.~\cite{Aaij:2015tna}:
\begin{align}
{\cal B}(B^0\to K^{*0}\chi){\cal B}(\chi\to\mu\mu) \lesssim
(0.8\mathchar`-6.3)\times 10^{-9},\; {\rm (LHCb)}
 \label{eq:B2KstrZp-LHCb}
\end{align}
at 95\% C.L. for 214 MeV $\leq m_{\chi} \leq$ 400 MeV, with $\tau_{\chi}\ll 1$ ps.
As the width of $Z'$ is very small, we neglect the interference between the $Z'$ and SM
contributions.

The LHCb result greatly improves the previous limit set by Belle~\cite{Hyun:2010an}:
\begin{align}
{\cal B}(B^0\to K^{*0}X){\cal B}(X\to\mu\mu)
\lesssim (2.3\mathchar`-5.0)\times 10^{-8},\ {\rm (Belle)}\label{eq:Belle}
\end{align}
at 90\% C.L. for a vector boson $X$ with mass in $212\,\MeV\leq m_{X} \leq 300\,\MeV$.
But the Belle result complements LHCb for the range $212\,\MeV\leq m_{X} \leq 214\,\MeV$ 
just above the dimuon threshold of $211.3$~MeV.

There are no existing results for the dedicated search of 
low-mass new bosons in the $B\to K\mu^+\mu^-$ mode.
We stress the importance of search in this mode, as the two decay modes
are complementary in probing the chiral structure of $bsZ'$ couplings:
the $B\to KZ'$ rate depends on the vector-like combination $g_{sb}^L+g_{sb}^R$,
while the $B\to K^*Z'$ rate is sensitive to the axial-vector combination
$g_{sb}^L-g_{sb}^R$, as can be read from Eq.~(\ref{eq:B2KZp-num}).

In a previous study~\cite{Fuyuto:2014cya}, published before the advent of
the LHCb analysis~\cite{Aaij:2015tna}, we attempted at constraining
the $Z'$ effect using existing LHCb data for $B^+\to K^+\mu^+\mu^-$.
We chose the 1 fb$^{-1}$ result~\cite{Aaij:2012vr} instead of
the 3 fb$^{-1}$ one~\cite{Aaij:2014pli}, as the latter provides the dimuon 
spectrum only for $q^2 > 0.1\,\GeV^2\simeq (316\,\MeV)^2$, which covers
only half the $Z'$ mass range in Scenario (ii-a).
The 1 fb$^{-1}$ result, however, gives the spectrum for $q^2 > 0.05\,\GeV^2$,
which can probe $m_{Z'}$ down to 224 MeV.
In contrast to $B\to K^*\mu^+\mu^-$, the photon peak is absent in $B^+\to K^+\mu^+\mu^-$, 
and the measured $q^2$ spectrum~\cite{Aaij:2012vr} is rather flat
in the low $q^2$ range, 
with average differential branching ratio 
$d{\cal B}/dq^2 = (2.41\pm 0.22)\times 10^{-8}\,\GeV^{-2}$ in $1\,\GeV^2 <q^2 <6\,\GeV^2$.
Treating this as background, we subtracted it from the measured value of 
$d{\cal B}/dq^2 = (2.85\pm 0.30)\times 10^{-8}\,\GeV^{-2}$
in the lowest $q^2$ bin of $0.05\,\GeV^2 < q^2 < 2.00\,\GeV^2$. 
We then estimated the allowed range of the $Z'$ contribution in this bin:
$\Delta {\cal B}(B^+\to K^+\mu^+\mu^-)=(0.86 \pm 0.59)\times 10^{-8}$.
At 2$\sigma$, this reads~\cite{Fuyuto:2014cya}
\begin{align}
\mathcal B(B^+\to K^+Z')\mathcal B(Z'\to\mu^+\mu^-) \lesssim
 2.0\times 10^{-8},~({\rm ``LHCb"}) \label{eq:B2KZp-LHCb}
\end{align}
for $224\,\MeV \lesssim m_{Z'} \lesssim 1414\,\MeV$.

\subsubsection{$B\to K^{(*)}\nu\bar\nu$ data}

In Scenario (ii-a), the other $\sim$50\% of $Z'$ bosons decay into neutrino pairs,
resulting in $B\to K^{(*)}\nu\bar\nu$.
Sensitivities of experimental searches for $B\to K^{(*)}\nu\bar \nu$ by
the BaBar~\cite{Lees:2013kla} and Belle~\cite{Lutz:2013ftz} experiments are still above the SM level.
For our purpose, the BaBar result~\cite{Lees:2013kla} is useful, as model-independent
constraints on BSM effects are given for spectra of $s_B\equiv m_{\nu\nu}^2/m_B^2$
bin by bin.
From Fig. 6 of Ref.~\cite{Lees:2013kla}, the first bin $0<s_B <0.1$, 
or $0<m_{\nu\nu}\lesssim 1670$ MeV, gives the constraints
$\Delta{\cal B}(B^+\to K^+\nu\bar\nu)=(0.35^{+0.60}_{-0.15})\times 10^{-5}$ and
$\Delta{\cal B}(B^+\to K^{*+}\nu\bar\nu)=(-0.1^{+1.9}_{-0.3})\times 10^{-5}$.
The other two decay modes with $K^0$ or $K^{*0}$ give weaker limits.
The $K^+$ channel favors nonzero BSM effects due to the observation of
a small excess over the expected background.
But the probability to observe such an excess in the signal region is 8.4\%,
hence is not significant.
The above limits at 2$\sigma$ imply, for $0<m_{Z'}\lesssim 1670$ MeV,
\begin{align}
&0.05 <10^5\, {\cal B}(B^+\to K^+Z'){\cal B}(Z'\to\nu\bar\nu) <1.55,
 \notag\\
&10^{5}\, {\cal B}(B^+\to K^{*+}Z'){\cal B}(Z'\to \nu\bar\nu)<3.7.\ \ ({\rm BaBar})
\label{eq:B2KZp-BaBar}
\end{align}

Given the $Z'$ branching ratios ${\cal B}_{\mu\mu}\sim {\cal B}_{\nu\nu}\sim 1/2$,
the excess in $B^+\to K^+\nu\bar\nu$ is not compatible with the LHCb limit on
$B^+\to K^+Z'(\to \mu^+\mu^-)$, Eq.~(\ref{eq:B2KZp-LHCb}). 
In Scenario (ii-a), we therefore treat the BaBar limits just as a reference,
except when the $Z'$ mass is close to the dimuon threshold and
the $B\to K^{(*)}\mu^+\mu^-$ limits do not apply.

\subsection{$t\to cZ'$ via left-handed current}

The $B\to K^*Z'$ rate is sensitive to the combination $g_{sb}^L-g_{sb}^R$, while its
dependence on $g_{sb}^L+g_{sb}^R$ is weaker, as can be seen from Eq.~(\ref{eq:B2KZp-num}).
Hence, the limits on $B^0\to K^{*0}Z'(\to\mu^+\mu^-)$ by LHCb, 
Eq.~(\ref{eq:B2KstrZp-LHCb}), draw an ellipse extending 
along the $g_{sb}^L = g_{sb}^R$ direction on the
($g_{sb}^{L},\,g_{sb}^R$) plane for each $m_{Z'}$ value.
The resulting constraints on the $bsZ'$ couplings are
$|g_{sb}^{L}|,\,|g_{sb}^R|\lesssim (2\mathchar`-7)\times 10^{-10}$
for $214\,\MeV \leq m_{Z'}\leq 400$ MeV.

The constraints on the $bsZ'$ couplings are improved if the limit on
$B^+\to K^+Z'(\to\mu^+\mu^-)$ extracted from the LHCb data, Eq.~(\ref{eq:B2KZp-LHCb}),
is further imposed: $|g_{sb}^{L}|,\,|g_{sb}^R|\lesssim 1\times 10^{-10}$ for $224\,\MeV
\lesssim m_{Z'}\leq 400$ MeV, which is rather stable w.r.t. $m_{Z'}$. 
The limit implies
\begin{align}
m_Q &\gtrsim 670~\TeV \sqrt{ |Y_{Qs}^*Y_{Qb}|
 \left(\frac{m_{Z'}}{300\,\MeV}\right)\left(\frac{10^{-3}}{g'}\right) },
\end{align}
which is an order of magnitude larger than in Scenario (i) [Eq.~(\ref{eq:b2s-sol-2})].
Assuming the SU(2)$_L$ relation $g_{ct}^L\simeq g_{sb}^L$, then, we obtain
bounds on the left-handed current contribution to the $t\to cZ'$ branching ratio:
\begin{align}
{\cal B}(t\to cZ')_{\rm LH} \lesssim (3\mathchar`-4)\times 10^{-15},
\end{align}
for $224\,\MeV \lesssim m_{Z'} \leq 400$ MeV.
These values would be too small to measure even with the high-luminosity LHC upgrade. 
[See Eq.~(\ref{eq:t2cZp-CMS}) for a naive expectation.]

For $214~\MeV \leq m_{Z'} \lesssim 224$ MeV, the $B^+\to K^+Z'(\to\mu^+\mu^-)$
limit does not apply, and bounds on the $t\to cZ'$ rate are weakened:
\begin{align}
{\cal B}(t\to cZ')_{\rm LH} \lesssim (2\mathchar`-4)\times 10^{-13},
\end{align}
for $214\,\MeV \leq m_{Z'} \lesssim 224$ MeV.
In the narrow interval $212\,\MeV \leq m_{Z'} < 214$ MeV, the LHCb limits on
$B^0\to K^{*0}Z'(\to\mu^+\mu^-)$ are taken over by the Belle limits,
Eq.~(\ref{eq:Belle}), which give an order of magnitude weaker bounds on
${\cal B}(t\to cZ')_{\rm LH}$ than the case for $214\,\MeV \leq m_{Z'} \lesssim 224$ MeV.
The remaining spot of 211.3 MeV $\lesssim m_{Z'} <$ 212 MeV, just above the
dimuon threshold, is still constrained by the BaBar limits on $B\to K^{(*)}\nu\bar\nu$,
Eq.~(\ref{eq:B2KZp-BaBar}), leading to
${\cal B}(t\to cZ')_{\rm LH} \lesssim 4\times 10^{-12}$, which is further diluted by
a small $Z'$ branching ratio ${\cal B}(Z'\to \mu^+\mu^-)\lesssim 10$\%.
Note that the excess in $B^+\to K^+\nu\bar\nu$ does not necessarily imply a nonzero
$g_{sb}^L\simeq g_{ct}^L$, as $g_{sb}^R$ alone can still explain the excess.
These values of ${\cal B}(t\to cZ')_{\rm LH}$ would be still too small for measurements
at the LHC.

In deriving the limits on the left-handed current contribution to ${\cal B}(t\to cZ')$,
we assumed the SU(2)$_L$ relation $g_{ct}^L\simeq g_{sb}^L$.
This is valid for $Y_{Qt}\sim Y_{Qc}$, but does not hold in general.
More precisely, the SU(2)$_L$ relation is given by Eq.~(\ref{eq:Y-SU(2)}):
$Y_{Qb}=V_{cb}Y_{Qc}+V_{tb}Y_{Qt}\simeq A\lambda^2 Y_{Qc}+Y_{Qt}$, and
$Y_{Qs}=V_{cs}Y_{Qc}+V_{ts}Y_{Qt}\simeq Y_{Qc}-A\lambda^2 Y_{Qt}$, 
where $A\simeq 0.81$~\cite{Agashe:2014kda}, 
and $Y_{Qu}$ is taken to be zero to avoid $D$ meson constraints.
A remarkable deviation from our assumption occurs when $Y_{Qc}/Y_{Qt}\sim \lambda^2$:
$Y_{Qs}$ vanishes for $Y_{Qc}\simeq A\lambda^2 Y_{Qt}$, hence,
$g_{sb}^L\propto Y_{Qs}^*Y_{Qb}\simeq 0$ (and $g_{ds}^L\simeq 0$).
This allows a large $g_{ct}^L$ without violating the $b\to sZ'$
(and $s\to dZ'$) constraints. Yet, this implies $Y_{Qd}\sim \lambda Y_{Qc}$ with
$Y_{Qb}\simeq Y_{Qt}$, hence, $g_{db}^L\propto Y_{Qd}^*Y_{Qb} \sim \lambda g_{ct}^L$,
which would be constrained by the measurement of the $B^+\to \pi^+\mu^+\mu^-$
decay by LHCb~\cite{LHCb:2012de,Aaij:2015nea}, as well as $B$-$\bar B$ mixing.
We do not pursue such an extreme case in this paper.

Before moving on, we briefly mention the $B_s$ meson mixing constraint.
For light $Z'$, the local $(\bar sb)(\bar sb)$ box operator construction 
in usual renormalization group analysis is not valid,
as the $Z'$ remains a dynamical degree of freedom at the $m_B$ scale.
Here, we simply recover the momentum dependence in the $Z'$ propagator
in the usual heavy $Z'$ formula and set the $Z'$ momentum to the $B_s$ mass scale.
To see the impact of this constraint, for simplicity,
we only include the left-handed $bsZ'$ coupling effects.
Employing the unitarity gauge and the vacuum insertion approximation~\cite{Oh:2009fm},
we find that the $B_s$-$\bar B_s$ mixing amplitude is modified as
\begin{align}
\frac{M_{12}}{M_{12}^{\rm SM}} &\simeq 1-\frac{(g_{sb}^L)^2v^2}{m_{B_s}^2-m_{Z'}^2}
 \left( 1-\frac{5}{8}\frac{m_{B_s}^2}{m_{Z'}^2} \right) \notag\\
&\quad \times \left[ \frac{g_2^2}{16\pi^2}(V_{ts}^*V_{tb})^2S_0 \right]^{-1}.
 \label{eq:B-mixing-light}
\end{align}
Allowing 15\% BSM effect, we obtain $|g_{sb}^L|\lesssim 2\times 10^{-6}
(m_{Z'}/300\,\MeV)$, four orders of magnitude weaker than
the constraint from $B\to K^{(*)}Z'(\to \mu^+\mu^-)$.

\subsection{Right-handed $tcZ'$ coupling: loop-induced down-quark sector FCNCs}

\begin{figure}[t!]
{
 \includegraphics[width=60mm]{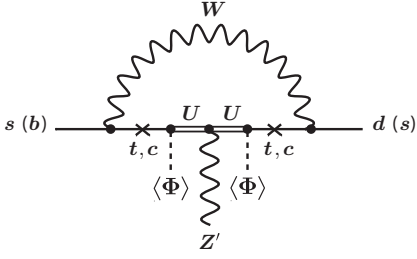}
}
\caption{
Feynman diagram that induces the effective $sdZ'$ ($bsZ'$) coupling at one-loop
through Yukawa couplings $Y_{Ut}$ and $Y_{Uc}$.
The crosses indicate quark mass insertions which flip chirality for $t$ or $c$.
We included similar contribution from the would-be Nambu-Goldstone bosons.
} \label{fig:sdZp}
\end{figure}

The right-handed $tcZ'$ coupling induces FCNCs in down-type quark sector at one-loop
level. More precisely, the SU(2)$_L$ singlet vector-like quark $U$, responsible for
the effective right-handed $tcZ'$ coupling, mediates the diagram in Fig.~\ref{fig:sdZp},
leading to extra contributions to effective $sdZ'$ and $bsZ'$ couplings.
Assisted by the SM charged current couplings of the $W$ boson to left-handed quarks,
not only $tcZ'$, but also flavor-diagonal $ttZ'$ and $ccZ'$ contribute.
These contributions are loop, chirality and CKM suppressed.
There are thus no significant impacts for the heavy $Z'$.
However, for light $Z'$, the loop-induced FCNC decays give meaningful
constraints, as the meson decay rates are hugely enhanced due to 
on-shell nature of the $Z'$, compensating these suppressions.

In estimating the loop-induced FCNC couplings, for simplicity,
we set $Y_{Qd_i}=Y_{Dd_i}=0$ ($i=1,2,3$) 
to turn off the tree-level FCNC couplings in the down-type quark sector.
We further set $Y_{Uu}=0$ to avoid constraints from $D$ meson decays and mixing,
for sake of maximizing the $t\to cZ'$ decay rate.
We are then left with the Yukawa mixing couplings for the vector-like quark $U$ 
to right-handed top or charm quarks, $Y_{Ut}$ and $Y_{Uc}$.

Working in the 't Hooft-Feynman gauge, we calculate the diagram of Fig.~\ref{fig:sdZp},
as well as similar contributions with the would-be Nambu-Goldstone bosons.
We neglect the external momenta as usual, but keep all internal momenta, including
the one for the vector-like quark $U$. We then obtain the loop-induced effective
couplings~\cite{Fuyuto:2014cya}:
\begin{align}
{\cal L}_{\rm eff} 
 \supset \Delta g_{d_id_j}^L
 \bar d_{iL}\gamma^\alpha d_{jL} Z'_\alpha +{\rm h.c.}, \label{eq:Leff-Z'-loop}
\end{align}
where $i,j=1,2,3$, and
\begin{align}
\Delta g_{d_id_j}^L & = \frac{g'v_\Phi^2}{32\pi^2v^2}
  \left[V_{td_j}V_{td_i}^* \kappa_{tt}f_{tt}
   + \left(V_{td_j}V_{cd_i}^* \kappa_{tc}\right.\right. \notag\\
& \quad \left.\left. + V_{cd_j}V_{td_i}^*\kappa_{ct}\right)f_{ct}
   + V_{cd_j}V_{cd_i}^* \kappa_{cc}f_{cc} \right],
 \label{eq:eff-coup-loop}
\end{align}
with
\begin{align}
\kappa_{u_ku_l} = Y_{Uu_k}Y_{Uu_l}^*\frac{m_{u_k}m_{u_l}}{m_U^2}, \label{eq:kappa}
\end{align}
and
\begin{align}
f_{tt} &\simeq \frac{3m_W^2}{m_t^2-m_W^2}\left( 1-\frac{m_W^2}{m_t^2-m_W^2}
          \log\frac{m_t^2}{m_W^2}\right) + \log\frac{m_U^2}{m_t^2}, \notag\\
f_{ct} &\simeq 1 + \log\frac{m_U^2}{m_t^2}
            + \frac{3m_W^2}{m_t^2-m_W^2}\log\frac{m_t^2}{m_W^2}, \notag \\
f_{cc} &\simeq 4\log\frac{m_W^2}{m_c^2} + \log\frac{m_U^2}{m_W^2}-3.
\label{eq:loop-func-apprx}
\end{align}
The expressions for these loop functions are in the large $m_U$ limit. 
Exact expressions used in our numerical study are given in Appendix \ref{app:loop}.


In the loop-induced $bsZ'$ and $bdZ'$ couplings, Eq.~(\ref{eq:eff-coup-loop}),
the top-top loop contribution dominates due to chiral factor,
so long as $Y_{Uc}/Y_{Ut}$ is not too large.
Then, $\Delta g_{db}^L/\Delta g_{sb}^L\simeq V_{td}^*/V_{ts}^* \sim \lambda$,
hence, the ratio ${\cal B}(b\to dZ')/{\cal B}(b\to sZ')\sim \lambda^2\simeq 0.05$ is SM-like.
For large $Y_{Uc}$ (e.g. $Y_{Uc}/Y_{Ut}\gtrsim 4$ for $m_U=2$ TeV),
the top-charm loop contribution becomes dominant.
Even in this case, $\Delta g_{db}^L/\Delta g_{sb}^L\simeq V_{cd}^*/V_{cs}^*\sim \lambda$,
hence, the ratio ${\cal B}(b\to dZ')/{\cal B}(b\to sZ')$ is still SM-like.
Thus, the better measured $b\to s$ decays are more suitable 
to watch $Z'$ effect than $b\to d$.
We consider $b\to sZ'$ and $s\to dZ'$ decays below.

\subsection{FCNC $K$ decays}

\subsubsection{$K\to\pi Z'$ formulas}

One can obtain ${\cal B}(B\to K^{(*)}Z')$ for loop-induced $bsZ'$ coupling 
by replacing $g_{sb}^L\to \Delta g_{sb}^L$, $g_{sb}^R\to 0$ in Eqs.~(\ref{eq:B2KZp})
and (\ref{eq:B2KsZp}).
Then, ${\cal B}(B\to K^{*}Z')\simeq {\cal B}(B\to KZ')$, as can be seen from
Eq.~(\ref{eq:B2KZp-num}). Hence, the LHCb limits on $B^0\to K^{*0}\chi(\to \mu^+\mu^-)$,
Eq.~(\ref{eq:B2KstrZp-LHCb}), give the strongest constraint in most $m_{Z'}$ ranges of
Scenario (ii-a).

The loop-induced $sdZ'$ coupling causes $K\to \pi Z'$ for $m_{Z'}<m_K-m_\pi$, 
leading to $K\to \pi \mu^+\mu^-$ and $K\to \pi \nu\bar\nu$ decays.
The branching ratios for $K^+\to\pi^+ Z'$ and $K_L \to\pi^0 Z'$ decays are given by
\begin{align}
&{\cal B}(K^+\to \pi^+ Z') \notag\\
&=\frac{|\Delta g_{ds}^L|^2}{64\pi}
 \frac{m_{K^+}^3}{m_{Z'}^2\Gamma_{K^+}}
 \beta^{3}_{K^+\pi^+Z'}\left[f_+^{K^+\pi^+}(m_{Z'}^2)\right]^2, \notag\\
&{\cal B}(K_L\to \pi^0 Z') \notag\\
&=\frac{[{\rm Im} (\Delta g_{ds}^L)]^2}{64\pi}
 \frac{m_{K_L}^3}{m_{Z'}^2\Gamma_{K_L}}
 \beta^{3}_{K_L\pi^0Z'}\left[f_+^{K^0\pi^0}(m_{Z'}^2)\right]^2, \label{eq:BR-K2piZp}
\end{align}
where $\beta_{K\pi Z'}$ is defined by Eq. (\ref{eq:beta-func}), and
$f_+^{K\pi}$ are the $K\to \pi$ form factors. For the latter, we adopt the result of
Ref.~\cite{Mescia:2007kn}, which is based on the partial NNLO calculation with
isospin-breaking effects in chiral perturbation theory.
In estimating the $K_L\to \pi^0Z'$ rate, we took
$| K_L \rangle \simeq (|K^0\rangle +|\bar K^0\rangle)/\sqrt{2}$ with the phase convention
where $CP|K^0\rangle =-|\bar K^0\rangle$, neglecting CP violation in kaon mixing.
The branching ratio for $K_S\to \pi^0Z'$ can be obtained from the one for $K_L\to \pi^0Z'$
with the replacements: ${\rm Im} (\Delta g_{ds}^L)\to {\rm Re} (\Delta g_{ds}^L)$ and
$\tau_{K_L}\to \tau_{K_S}$.

\subsubsection{$K\to \pi\mu^+\mu^-$ data}

In SM, the $K^+\to\pi^+\mu^+\mu^-$ decay is dominated by
long-distance effects via one-photon exchange $K^+\to\pi^+\gamma^*$.
The decay can be described by chiral perturbation theory~\cite{Ecker:1987qi}
with the dimuon invariant mass spectrum $d\Gamma/dz\propto |W(z)|^2$, where
$z\equiv m_{\mu\mu}^2/m_{K^+}^2$ and $W(z)$ is the $K\to\pi\gamma^*$ form factor.
The most precise value for ${\cal B}(K^+\to\pi^+\mu^+\mu^-)$ by a single measurement
comes from the NA48/2 experiment~\cite{Batley:2011zz} at the CERN SPS.
The measured $z$-spectrum in the whole kinematic range of
$4m_\mu^2/m_{K^+}^2 \leq z \leq (1-m_{\pi^+}/m_{K^+})^2$,
corresponding to 211 MeV $\lesssim m_{\mu\mu}\lesssim$ 354 MeV,
is reasonably described by various form factor models.
In particular, the measured $z$-spectrum does not exhibit significant excesses over
the fit curve by the linear form factor model.
Here, we attempt a simple data-based approach to extract reasonable sizes for
possible $Z'$ effects.

In a previous study~\cite{Fuyuto:2014cya}, we focused on the largest upward deviation
from the fit curve in the $z$-spectrum of NA48/2, which is located in
the $z\in (0.32,0.34)$ bin, corresponding to $m_{\mu\mu}\in (279, 288)$ MeV.
Subtracting the fit value from the measured one, we read the allowed range for
an extra contribution: $\Delta(d\Gamma/dz)\simeq (2.5\pm 1.5)\times 10^{-24}$ GeV.
This corresponds to the deviation of the branching ratio in $m_{\mu\mu}\in (279, 288)$ MeV:
$\Delta\mathcal B(K^+\to \pi^+\mu^+\mu^-) \simeq (9.4\pm 5.6) \times 10^{-10}$.
Allowing $2\sigma$ range, we estimate the limit on the $Z'$ contribution:
${\cal B}(K^+\to \pi^+ Z'){\cal B}(Z'\to \mu^+\mu^-) \lesssim 2.1 \times 10^{-9}$
for 279 MeV $\lesssim m_{Z'} \lesssim$ 288 MeV.
The constraint is tighter for other $Z'$ mass values in
211 MeV $\lesssim m_{\mu\mu}\lesssim$ 354 MeV. For instance, we obtain
\begin{align}
{\cal B}(K^+\to \pi^+ Z'){\cal B}(Z'\to \mu\mu) \lesssim 1.1\times 10^{-9}~({\rm ``NA48/2"})
 \label{eq:NA48-334}
\end{align}
for 327 MeV $< m_{Z'} \lesssim$ 335 MeV, and
\begin{align}
{\cal B}(K^+\to \pi^+ Z'){\cal B}(Z'\to \mu\mu) \lesssim 1.2\times 10^{-9}~({\rm ``NA48/2"})
 \label{eq:NA48-219}
\end{align}
for $2m_\mu < m_{Z'} \lesssim$ 221 MeV.

For $K_L\to \pi^0\mu^+\mu^-$,
the current best limit comes from KTeV~\cite{AlaviHarati:2000hs} at Fermilab,
giving the 90\% C.L. limit
\begin{align}
{\cal B}(K_L\to \pi^0\mu^+\mu^-)<3.8\times 10^{-10}.\ ({\rm KTeV})
\label{eq:KTeV}
\end{align}
This is above the SM prediction of 
$(1.29^{+0.24}_{-0.23})\times 10^{-11}$~\cite{Mertens:2011ts}.
We thus impose the above limit on ${\cal B}(K_L\to \pi^0Z'){\cal B}(Z'\to\mu^+\mu^-)$
for $2m_\mu < m_{Z'}<$ 350 MeV, covered by the kinematic selection of the KTeV analysis.

The $K_S\to \pi^0\mu^+\mu^-$ mode was measured by NA48/1~\cite{Batley:2004wg} 
at CERN SPS, giving ${\cal B}(K_S\to \pi^0\mu^+\mu^-)
=[2.9^{+1.5}_{-1.2}({\rm stat})\pm 0.2({\rm syst})]\times 10^{-9}$.
This was used as input in the SM prediction of $K_L\to \pi^0\mu^+\mu^-$~\cite{Mertens:2011ts} 
to control the indirect CP violating contribution.
For the possible $Z'$ contribution, isospin symmetry implies
${\cal B}(K_S\to \pi^0 Z')\lesssim (\tau_{K_S}/\tau_{K^+}){\cal B}(K^+\to\pi^0Z')
\simeq 0.007\times{\cal B}(K^+\to\pi^0Z')$. Given that the experimental sensitivity on
the $K^+\to \pi^+Z'(\to \mu^+\mu^-)$ branching ratio is around $10^{-9}$, the above
isospin relation constrains the $K_S\to \pi^0 Z'(\to \mu^+\mu^-)$ branching ratio to be
within $\sim 10^{-11}$, which may be beyond the sensitivity of NA48/1 data.

\subsubsection{$K\to \pi\nu\bar\nu$ data}

For $K^+\to \pi^+\nu\bar\nu$ decay, the E949 experiment~\cite{Artamonov:2008qb}
at BNL, together with its predecessor E787, reported 
${\cal B}(K^+\to \pi^+\nu\bar\nu)=(1.73^{+1.15}_{-1.05}) \times 10^{-10}$,
which is consistent with SM prediction of $(8.25\pm 0.64)\times 10^{-11}$~\cite{Mertens:2011ts}.
The measurement error is large and E787/E949~\cite{Artamonov:2009sz}
also reported the 90\% C.L. upper limit of
${\cal B}(K^+\to \pi^+\nu\bar\nu)<3.35 \times 10^{-10}$.
We remark, however, that the experimental analyses utilized limited intervals
for the pion momentum $p_{\pi^+}$, or 
equivalently the neutrino pair mass $m_{\nu\nu}$,
to avoid blinding backgrounds from $K^+\to \pi^+\pi^0$ and
$K^+\to \pi^+\pi^-\pi^+/\pi^+\pi^0\pi^0$:
one is the $\pi\nu\bar\nu(1)$ region, where 211 MeV $<p_{\pi^+} <$ 229 MeV,
or $0\leq m_{\nu\nu} \lesssim$ 116 MeV;
the other is the $\pi\nu\bar\nu(2)$ region, where 140 MeV $<p_{\pi^+} <$ 199 MeV,
or 152 MeV $\lesssim m_{\nu\nu} \lesssim$ 261 MeV.
The kinematic selection of the $K^+\to \pi^+\nu\bar\nu$ experiments has an interesting
implication for $K_L\to\pi^0\nu\bar\nu$ search~\cite{Fuyuto:2014cya},
as we will discuss in the next section.

The E787/E949 data have been used also for a dedicated search \cite{Artamonov:2009sz}
of a two-body decay $K^+ \to \pi^+ P^0$ with $P^0\to\nu\bar\nu$,
where $P^0$ is a hypothetical short-lived particle.
The upper limits on ${\cal B}(K^+\to \pi^+ P^0){\cal B}(P^0\to\nu\bar\nu)$ were given
for the mass ranges of $0\leq m_{P^0} \lesssim 125$ MeV or 150 MeV
$\lesssim m_{P^0} \lesssim$ 260 MeV, which correspond to $\pi\nu\bar\nu(1)$ or
$\pi\nu\bar\nu(2)$ regions, respectively.
In the mass range which is relevant to Scenario (ii-a), the 90\% C.L. upper limits
increases almost monotonically with mass within
\begin{align}
{\cal B}(K^+\to \pi^+ P^0){\cal B}(P^0\to\nu\bar\nu) \lesssim
(0.4\mathchar`-5)\times 10^{-9},\ {\rm (E949)} \label{eq:E949-260}
\end{align}
for $2m_\mu < m_{P^0} \lesssim$ 260 MeV.

To facilitate the discussion in Scenario (ii-b), we also quote 90\% C.L. upper limits for
typical $P^0$ masses below the dimuon threshold by setting ${\cal B}(P^0\to\nu\bar\nu)=1$.
For $0\leq m_{P^0} \lesssim 125$ MeV,
the strongest (weakest) bound is attained for $m_{P^0}\simeq 95$ (125) MeV with
${\cal B}(K^+\to \pi^+ P^0) \lesssim 5\times 10^{-11}~(4\times 10^{-9})$.
The bound is rather stable for $0\leq m_{P^0} \lesssim 40$ MeV with
${\cal B}(K^+\to \pi^+ P^0) \lesssim 10^{-10}$.
For 150 MeV $\lesssim m_{P^0} <2m_\mu$,
the strongest (weakest) bound is attained for $m_{P^0}\simeq 190$ (150) MeV with
${\cal B}(K^+\to \pi^+ P^0)\lesssim 4\times 10^{-10}~(10^{-8})$.

\begin{table*}[t!]
{
$$
\begin{tabular}{l|l|ccc|l}
\hline\hline
Mode & Experiment & $m_{Z'}=334$ MeV & $m_{Z'}=219$ MeV & $m_{Z'}=135$ MeV
 & Comment\\
\hline
$B^0\to K^{*0}Z' (\to \mu^+\mu^-)$ & LHCb \cite{Aaij:2015tna} & $<4.41\times 10^{-9}$
 & $<6.29\times 10^{-9}$ & -- & See Eq. (\ref{eq:LHCb-219-334}). \\
$B^+\to K^+Z'(\to \nu\bar{\nu})$ & BaBar~\cite{Lees:2013kla} &
 $(0.05,1.55)\times 10^{-5}$ & $(0.05,1.55)\times 10^{-5}$ & $(0.05,1.55)\times 10^{-5}$
 & See Eq. (\ref{eq:B2KZp-BaBar}).\\
\hline
$K^+\to \pi^+ Z'(\to \mu^+\mu^-)$ & ``NA48/2'' \cite{Batley:2011zz} &
 $\lesssim 1.1\times 10^{-9}$ & $\lesssim 1.2\times 10^{-9}$ & --
 & See Eqs. (\ref{eq:NA48-334}), (\ref{eq:NA48-219}).\\
$K_L\to \pi^0 Z'(\to \mu^+\mu^-)$ & KTeV \cite{AlaviHarati:2000hs} &
 $<3.8\times 10^{-10}$ & $<3.8\times 10^{-10}$ & --
 & See Eq. (\ref{eq:KTeV}). \\
\hline
$K^+\to \pi^+ Z'(\to \nu\bar\nu)$ & E787/E949 \cite{Artamonov:2009sz,Artamonov:2005cu}
 & -- & $\lesssim 5\times 10^{-10}$ & $< 5.6\times 10^{-8}$
 & See Eqs. (\ref{eq:E949-260}), (\ref{eq:E949-pi0}). \\
$K_L\to \pi^0 Z'(\to \nu\bar{\nu})$ & E391a \cite{Ahn:2009gb} & $<2.6\times 10^{-8}$
 & $<2.6\times 10^{-8}$ & $<2.6\times 10^{-8}$
 & See Eq. (\ref{eq:E391a}). \\
\hline
\hline
\end{tabular}
$$
}
\caption[]{
Summary of $B$ and $K$ decay constraints for the three benchmark points
in scenarios (ii-a) and (ii-b): $m_{Z'}=334$, $219$, $135$ MeV.
The numbers shown in third to fifth column are the allowed ranges for
each branching ratios, used in Figs.~\ref{fig:219-334_Y}--\ref{fig:135_mU}.
See text, and in particular the referred equations (last column), for detail.
}\label{tab:exp}
\end{table*}

For the pocket $125~\MeV \lesssim m_{Z'}\lesssim 150$ MeV around $\pi^0$ mass,
the upper limit can be still obtained by using the $\pi^0\to \nu\bar\nu$ 
search in $K^+\to \pi^+\pi^0$ by E949~\cite{Artamonov:2005cu}:
${\cal B}(\pi^0\to\nu\bar\nu)<2.7\times 10^{-7}$ at 90\% C.L.
In this search, charged pions with momentum in $198\,\MeV <p_{\pi^+}<212$ MeV were selected, 
corresponding to $112\,\MeV \lesssim m_{\nu\nu} \lesssim 155$ MeV,
hence the $\pi^0$-pocket can be fully covered.
Combining with ${\cal B}(K^+\to \pi^+\pi^0)\simeq 20.7$\% \cite{Agashe:2014kda}, 
one has
\begin{equation}
{\cal B}(K^+ \to \pi^+ Z') < 5.6 \times 10^{-8}, \quad ({\rm E949}) \label{eq:E949-pi0}
\end{equation}
at 90\% C.L. for $112 \lesssim m_{Z'}\lesssim 155$ MeV.

The $K_L\to\pi^0\nu\bar\nu$ decay has been searched for
by the E391a experiment~\cite{Ahn:2009gb} at the KEK proton synchrotron,
setting the 90\% C.L. upper limit
\begin{align}
\mathcal B(K_L\to\pi^0\nu\bar\nu) < 2.6\times 10^{-8}, \quad ({\rm E391a}) \label{eq:E391a}
\end{align}
without any particular cut on $m_{\nu\nu}$.
This is far above the SM prediction of  $(2.60\pm 0.37)\times 10^{-11}$
\cite{Mertens:2011ts}.
Therefore, we impose Eq.~(\ref{eq:E391a}) on ${\cal B}(K_L\to \pi^0Z')
{\cal B}(Z'\to \nu\bar\nu)$ for $m_{Z'}< m_{K_L}-m_{\pi^0}\simeq 363$ MeV.
Note that in Scenario (ii-a), where $m_{Z'}>2m_\mu$, the KTeV limit, Eq.~(\ref{eq:KTeV}),
gives stronger constraint in general, as ${\cal B}_{\mu\mu}\sim {\cal B}_{\nu\nu} \sim 1/2$.
There are no existing constraints on $K_S\to\pi^0\nu\bar\nu$, where
the branching ratio is suppressed by $\tau_{K_S}/\tau_{K_L}\ll 1$ compared to
$K_L\to\pi^0\nu\bar\nu$.


%
\subsection{$t\to cZ'$ via right-handed current}

We can now combine all $B$ and $K$ decay data to constrain the right-handed current 
contribution ${\cal B}(t\to cZ')_{\rm RH}$. 
For illustration, we take the two benchmark points
(shown by red crosses in Fig.~\ref{fig:lifetime}):
\begin{itemize}
\item $m_{Z'}=334$ MeV, $g'=1.4\times 10^{-3}$, $m_U=2$ TeV,
\item $m_{Z'}=219$ MeV, $g'=1.1\times 10^{-3}$, $m_U=2$ TeV.
\end{itemize}
The $Z'$ mass values are chosen such that the LHCb limit for
$B^0\to K^{*0}\chi(\to\mu^+\mu^-)$, Eq.~(\ref{eq:B2KstrZp-LHCb}), is weakened to be
tolerant for a possible large $t\to cZ'$ rate:
$m_{Z'}=219$ MeV is one of the points which give the weakest limit in the {\it whole} range
of $214~\MeV \leq m_{Z'} \leq 400$ MeV, while
$m_{Z'}=334$ MeV gives the weakest limit in
the {\it high} mass region $260~\MeV \lesssim m_{Z'}\leq 400$ MeV, 
where the E949 limits for $K^+\to \pi^+P^0(\to \nu\bar\nu)$ do not apply.
The two benchmark points are phenomenological representatives of
$B$ and $K$ decay constraints.
The 95\% C.L. upper limits by LHCb \cite{Aaij:2015tna} are~\footnote{
 We thank M. Williams for providing us the precise upper values used in our study.
}
\begin{align}
&{\cal B}(B^0\to K^{*0}\chi){\cal B}(\chi\to\mu^+\mu^-) \notag \\
&< \begin{cases}
 4.41 \times 10^{-9} & (m_{\chi}=334~\MeV),\\
 6.29 \times 10^{-9} & (m_{\chi}=219~\MeV).
\end{cases} \label{eq:LHCb-219-334}
\end{align}
We neglect again the changes in efficiencies from scalar boson case.
For loop-induced $bsZ'$ coupling, where $g_{sb}^R=0$,
the changes are indeed extremely small, up to 6\% in the mass range
of our interest (See Appendix~\ref{app:B2Kpimumu}).
The $B$ and $K$ constraints are summarized in Table~\ref{tab:exp}.

\begin{figure*}[t!]
{\includegraphics[width=75mm]{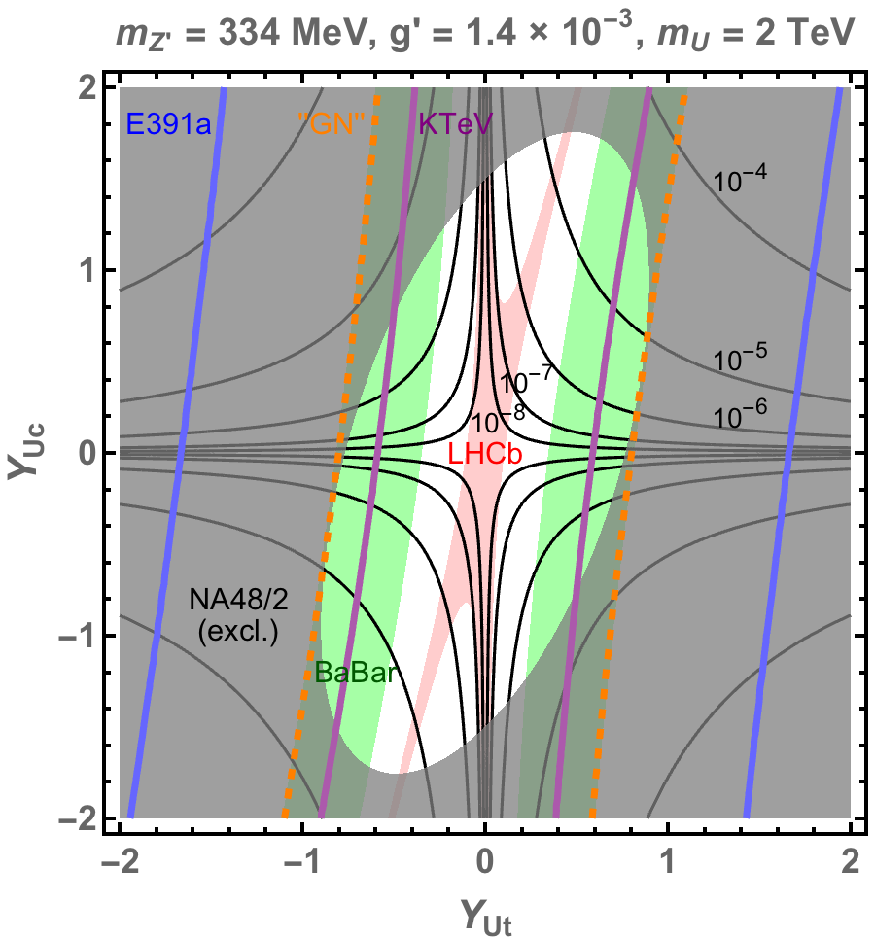} \hskip0.5cm
 \includegraphics[width=78mm]{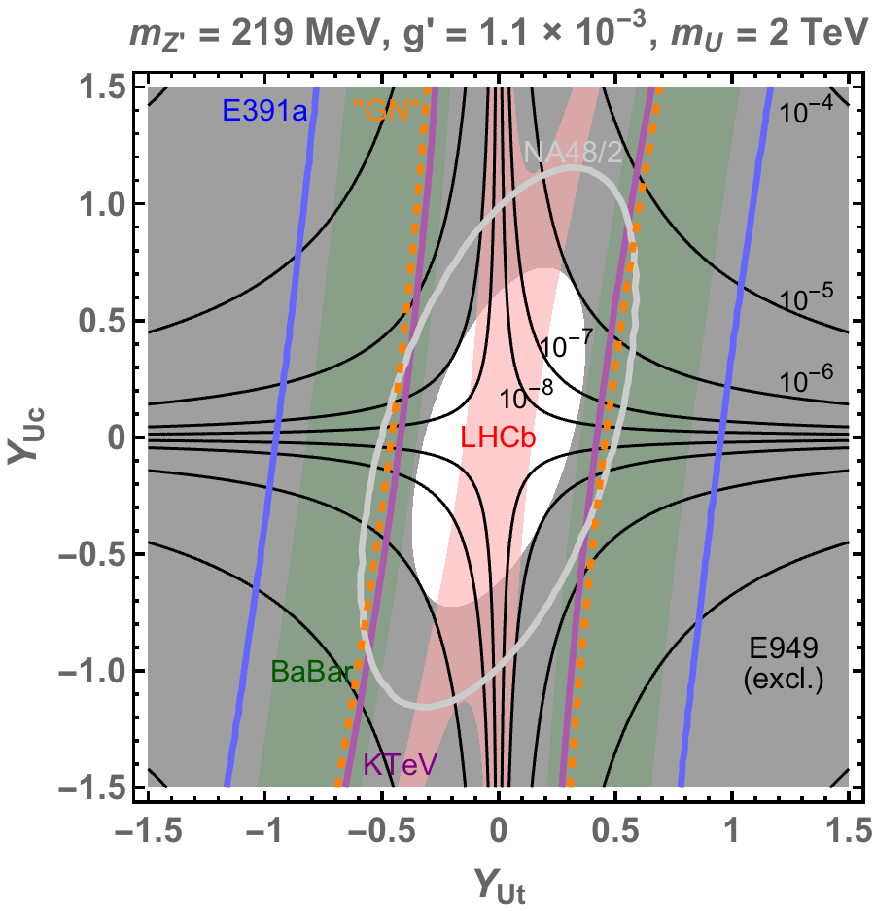}
}
\vskip0.05cm
\caption{
[Left] Contours of ${\cal B}(t\to cZ')_{\rm RH}$ are given as black-solid lines in
the $(Y_{Ut},\,Y_{Uc})$ plane for $m_{Z'}=334$ MeV, $g'= 1.4\times 10^{-3}$ and $m_U=2$ TeV.
The pink shaded region is allowed by the LHCb 95\% C.L. limit for
$B^{0}\to K^{*0}\chi(\to \mu^+\mu^-)$ in Eq.~(\ref{eq:LHCb-219-334}).
The light-green shaded regions are favored by the mild excess in BaBar data for
$B^+\to K^+\nu\bar\nu$ at $2\sigma$ [Eq.~(\ref{eq:B2KZp-BaBar})].
The semi-transparent dark-gray shaded region represents 2$\sigma$ exclusion by
NA48/2 for $K^+\to\pi^+\mu^+\mu^-$, Eq.~(\ref{eq:NA48-334}).
The purple-solid lines are 90\% C.L. exclusion by KTeV for $K_L\to\pi^0\mu^+\mu^-$,
Eq. (\ref{eq:KTeV}). The blue-solid lines are 90\% C.L. exclusion by E391a for $K_L\to\pi^0\nu\bar\nu$, Eq. (\ref{eq:E391a}). The orange-dashed lines are
the usual ``GN bound'' of Eq. (\ref{eq:usual-GN}), explained later.
[Right] Same as left panel, but for $m_{Z'}=219$ MeV, $g'= 1.1\times 10^{-3}$ and
$m_U=2$ TeV.
Here, the NA48/2 exclusion, Eq. (\ref{eq:NA48-219}), is shown by the light-gray solid line,
while the semi-transparent dark-gray-shaded region is excluded by the E949 90\% C.L. limit
${\cal B}(K^+\to\pi^+ P^0){\cal B}(P^0\to\nu\bar\nu)\lesssim 5\times 10^{-10}$.
} \label{fig:219-334_Y}
\end{figure*}

In Fig.~\ref{fig:219-334_Y} [left], we give contours of ${\cal B}(t\to cZ')_{\rm RH}$ as
black-solid lines in the $(Y_{Ut},\,Y_{Uc})$ plane for $m_{Z'}=334$ MeV benchmark point.
The meson decay constraints are imposed by taking into account the $Z'$ branching ratios:
${\cal B}_{\mu\mu}\simeq 48$\%, ${\cal B}_{\nu\nu}\simeq 52$\%.
The pink-shaded region is allowed by the LHCb bound on
$B^{0}\to K^{*0}\chi(\to \mu^+\mu^-)$ in Eq.~(\ref{eq:LHCb-219-334}).
The light-green-shaded regions are favored by the mild excess in BaBar data for
$B^+\to K^+\nu\bar\nu$ at 2$\sigma$ [Eq.~(\ref{eq:B2KZp-BaBar})].
The semi-transparent dark-gray shaded region represents 2$\sigma$ exclusion by
the NA48/2 data for $K^+\to\pi^+\mu^+\mu^-$, Eq.~(\ref{eq:NA48-334}),
from our illustration of the limit on the $Z'$ effect.
The purple-solid lines are 90\% C.L. exclusion by KTeV data for $K_L\to\pi^0\mu^+\mu^-$,
Eq.~(\ref{eq:KTeV}).

All the data have better sensitivity for $Y_{Ut}$ than $Y_{Uc}$, as the top-top
loop contribution dominates the loop-induced effective couplings,
Eq.~(\ref{eq:eff-coup-loop}), due to $m_t/m_c$ chiral enhancement.
The LHCb limit for $B^0\to K^{*0}\chi(\to \mu^+\mu^-)$ provides the strongest
constraint, 
excluding the BaBar region that could account for the $B^+\to K^+\nu\bar\nu$ excess.
Nevertheless, the LHCb limit accommodates ${\cal B}(t\to cZ')_{\rm RH}\gtrsim 10^{-6}$
along the funnel regions, extending towards large $Y_{Uc}$.
However, the NA48/2 limit for $K^+\to\pi^+\mu^+\mu^-$ eventually 
cuts down these funnels.
We obtain ${\cal B}(t\to cZ')_{\rm RH}\lesssim 2\times 10^{-5}$.

We remark that $m_{Z'}=334$ MeV is close to the kinematical limit 
$m_{K^+}-m_{\pi^+}\simeq 354$ MeV of $K^+\to \pi^+Z'$, 
and the NA48/2 data is less constraining than generic cases.
This can be seen from the {\it velocity} factor in Eq.~(\ref{eq:BR-K2piZp}):
$\beta_{K^+\pi^+Z'}\simeq 0.26$ for $m_{Z'}=334$ MeV, leading to the suppression of
${\cal B}(K^+\to \pi^+Z')$ by $\beta_{K^+\pi^+Z'}^3\simeq 0.018$, compared with,
e.g. $\beta_{K^+\pi^+Z'}^3\simeq 0.31\,(0.080)$ for $m_{Z'}=219\,(300)$ MeV.

For $m_{Z'}> m_{K^+}-m_{\pi^+}$, the two-body decay $K^+\to \pi^+Z'$
is kinematically forbidden, and the NA48/2 data loses constraining power,
hence ${\cal B}(t\to cZ')_{\rm RH}$ can be arbitrary large along the funnels.
However, the funnels imply some degree of fine-tuning between $Y_{Ut}$ and
$Y_{Uc}$ with cancelled contributions to $b\to sZ'$.
Furthermore, $Y_{Uc}$ should not be too large to maintain perturbativity.

A similar plot for the $m_{Z'}=219$ MeV benchmark point is given in
Fig.~\ref{fig:219-334_Y} [right],
where ${\cal B}_{\mu\mu}\simeq 28$\%, ${\cal B}_{\nu\nu}\simeq 72$\%.
In this case, the E949 limit for $K^+\to\pi^+ P^0(\to \nu\bar\nu)$ enters:
${\cal B}(K^+\to\pi^+ P^0){\cal B}(P^0\to\nu\bar\nu)
 \lesssim 5\times 10^{-10}$ at 90\% C.L.~\cite{Artamonov:2009sz}, 
shown as semi-transparent dark-gray shaded exclusion region.
This surpasses the NA48/2 limit for $K^+\to\pi^+\mu^+\mu^-$,
Eq.~(\ref{eq:NA48-219}), shown by the light-gray solid ellipse in the figure.
The E949 limit fully excludes the funnel regions, and we obtain
${\cal B}(t\to cZ')_{\rm RH}\lesssim 0.8\times 10^{-6}$.

The E949 limit gets stronger towards $m_{Z'}=2m_\mu$ and generically excludes
the funnel regions for $2m_\mu < m_{Z'}\lesssim 230$ MeV,
leading to ${\cal B}(t\to cZ')_{\rm RH}\lesssim 10^{-6}$.
Remarkably, this limit on ${\cal B}(t\to cZ')_{\rm RH}$ holds even in the pocket
$211.3~\MeV \lesssim m_{Z'} < 212$ MeV, where neither LHCb nor Belle limits
for $B^0\to K^{*0}Z'(\to \mu^+\mu^-)$ apply.

We have used $m_U=2$ TeV, but obtained similar results for other $m_U$ values.
This is because both $g_{ct}^R$ and $\Delta g_{sb(ds)}^L$ are proportional to $m_U^{-2}$,
up to logarithmic dependence, multiplied by a quadratic form in $Y_{Ut}$ and $Y_{Uc}$
[See Eqs. (\ref{eq:eff-coup}) and (\ref{eq:eff-coup-loop})].
Thus, changing $m_U$ simply results in rescaled $Y_{Ut}$ and $Y_{Uc}$ values.
A similar argument applies to the dependence on the U(1)$'$ coupling $g'$.

\begin{figure*}[t!]
{\includegraphics[width=75mm]{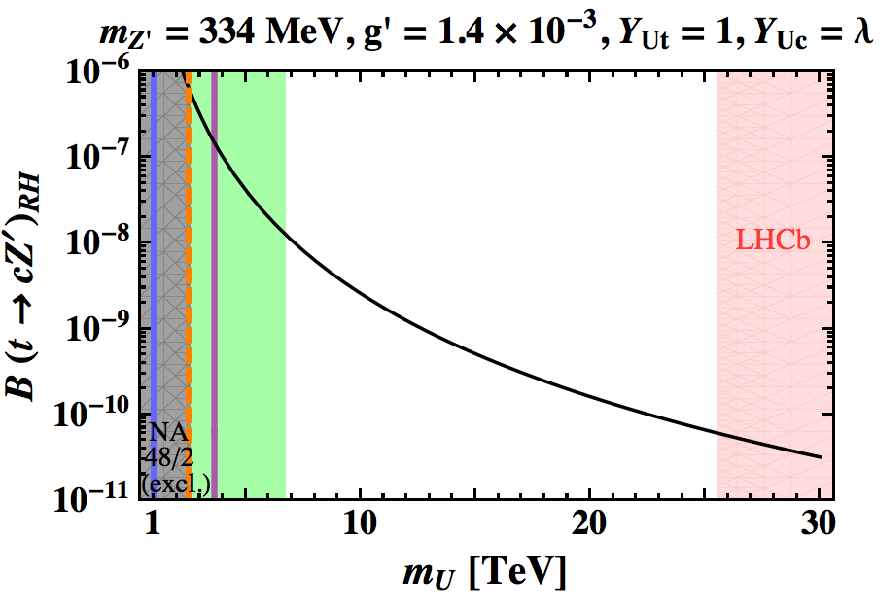} \hskip0.5cm
 \includegraphics[width=75mm]{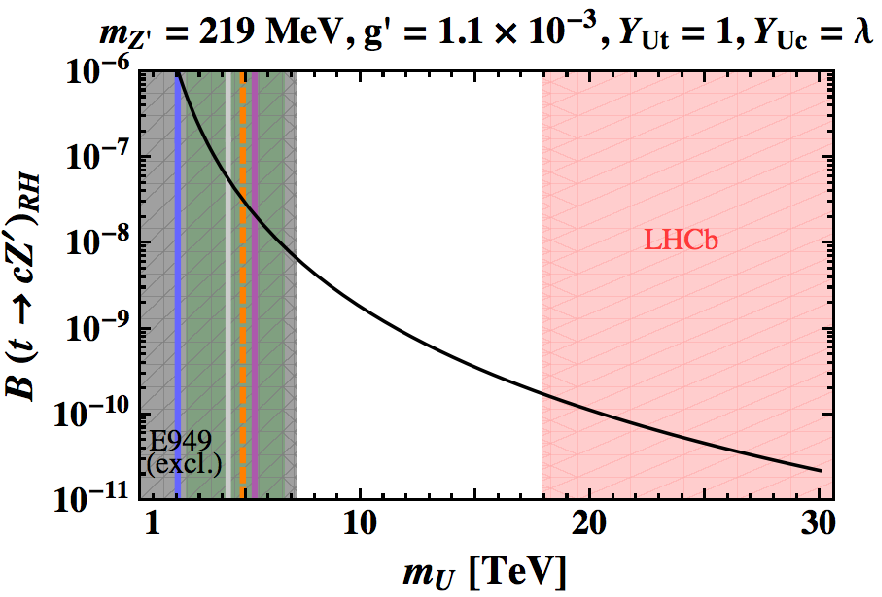}
}
\vskip0.05cm
\caption{
[Left]
Branching ratio ${\cal B}(t\to cZ')_{\rm RH}$ of right-handed current mediated 
$t\to cZ'$ as a function of vector-like quark mass $m_U$ for $m_{Z'}=334$ MeV
and $g'= 1.4\times 10^{-3}$ with hierarchical Yukawa couplings
 $Y_{Ut}=1$, $Y_{Uc}=\lambda\simeq 0.23$.
Shaded regions and lines are constraints on $m_U$ from $B$ and $K$ decay data,
with the shadings and lines as explained in Fig.~\ref{fig:219-334_Y}.
[Right]
Same as left panel, but for $m_{Z'}=219$ MeV and $g'= 1.1\times 10^{-3}$.
} \label{fig:219-334_mU}
\end{figure*}

As we took $Y_{Uu}=0$ to avoid $D$ meson constraints, one might think that
$Y_{Ut} > Y_{Uc} > Y_{Uu}$ is the natural ordering of these Yukawa couplings.
Taking $Y_{Ut}=1$ and $Y_{Uc}=\lambda$, we plot in Fig.~\ref{fig:219-334_mU} [left]
${\cal B}(t\to cZ')_{\rm RH}$ as a function of $m_U$ for $m_{Z'}=334$ MeV,
with $B$ and $K$ constraints overlaid.
In this case, the LHCb constraint is severe, implying
${\cal B}(t\to cZ')_{\rm RH}\lesssim 0.6\times 10^{-10}$.
A similar plot for $m_{Z'}=219$ MeV is given in Fig.~\ref{fig:219-334_mU} [right],
where the LHCb limit implies
${\cal B}(t\to cZ')_{\rm RH}\lesssim 2\times 10^{-10}$.
The pocket $211.3~\MeV \lesssim m_{Z'} < 212$ MeV is still constrained by E949,
giving ${\cal B}(t\to cZ')_{\rm RH}\lesssim 3\times 10^{-9}$.

In short, for the hierarchical Yukawa couplings $Y_{Ut}=1$, $Y_{Uc}=\lambda$,
we obtain ${\cal B}(t\to cZ')_{\rm RH}\lesssim {\cal O}(10^{-9})$ in the whole mass range
of Scenario (ii-a), namely $2m_\mu < m_{Z'}\lesssim 400$ MeV.
These values seem beyond reach of the LHC.
On the other hand, if $Y_{Ut}$ and $Y_{Uc}$ are treated as free parameters,
${\cal B}(t\to cZ')_{\rm RH}$ can be much larger.
In particular, ${\cal B}(t\to cZ')_{\rm RH}\sim {\cal O}(10^{-5})$ is possible in the funnel 
regions for appropriately high $Z'$ masses such that the E949 limit is weakened.
This may be within reach at the future LHC, as shown in Eq.~(\ref{eq:t2cZp-CMS}), 
but a fine-tuned correlation between $Y_{Ut}$ and $Y_{Uc}$ would be needed.
Note that our projection for LHC sensitivities is rather naive.
A careful collider study should be done to judge the 
actual sensitivity for $t\to cZ'$ at the LHC.


Surveying other $m_{Z'}$ cases, we find the constraints on ${\cal B}(t\to cZ')_{\rm RH}$ in Scenario (ii-a) can be classified into the following three categories.
\begin{itemize}
\item $330~\MeV \lesssim m_{Z'}\lesssim 400$ MeV: \\
 ${\cal B}(t\to cZ')_{\rm RH}\gtrsim 10^{-6}$ is possible along
 the funnel regions allowed by hidden-sector boson search of LHCb in
 $B^0\to K^{*0}\chi(\to \mu^+\mu^-)$. In particular, if $m_{Z'}\gtrsim m_{K^+}-m_{\pi^+}
 \simeq 354$ MeV, the $K^+$ decay constraints can be avoided, and
 ${\cal B}(t\to cZ')_{\rm RH}$ may be arbitrary large for large $Y_{Uc}$,
 up to perturbativity and associated fine-tuning of $Y_{Ut}$.
\item $230~\MeV \lesssim m_{Z'}\lesssim 330$ MeV: \\
 ${\cal B}(t\to cZ')_{\rm RH}\gtrsim 10^{-6}$ is possible along the funnel regions,
 but $K^+\to \pi^+\mu^+\mu^-$ (NA48/2) and/or $K^+\to\pi^+P^0(\to\nu\bar\nu)$ (E949)
 constraints cut in, such that ${\cal B}(t\to cZ')_{\rm RH}\lesssim 10^{-5}$.
\item $2m_\mu < m_{Z'}\lesssim 230$ MeV: \\
 The E949 limit gets stronger towards $m_{Z'}=2m_\mu$, fully excluding the funnel regions, 
 giving ${\cal B}(t\to cZ')_{\rm RH}\lesssim 10^{-6}$, even in the
 $211.3~\MeV \lesssim m_{Z'} < 212$ MeV pocket, where neither LHCb nor Belle limits
 for $B^0\to K^{*0}Z'(\to \mu^+\mu^-)$ apply.
\end{itemize}

We have ignored the interference of $Z'$ effect with SM contribution, 
as the $Z'$ width is tiny.
We remark, however, that for $m_{Z'}>2m_\pi$, 
an absorptive part of $K^+\to\pi^+\gamma^*(\to\mu^+\mu^-)$ induced by the $\pi\pi$ loop 
may invalidate the simple separation of SM and $Z'$ contributions to $K^+\to \pi^+\mu^+\mu^-$.
This might affect the second mass range, in particular
the position where the funnels are cut down by the NA48/2 limit.

\section{A light $Z'$ that evades Grossman-Nir bound \label{sec:light-nunu}}

In this section, we study the scenario where
\begin{align}
m_{Z'}< 2m_\mu  \quad [{\rm Scenario\,(ii\mathchar`-b)}].
\end{align}
In this case, the $Z'$ bosons decay exclusively into neutrino pairs
and are just felt as missing energy in collider experiments.
As such, it would be more challenging to search for $t\to cZ'$ at the LHC.
Nevertheless, we estimate for completeness the allowed ranges for 
$t\to cZ'$ branching ratios via left- or right-handed current in this scenario.
The relevant formulas and meson decay constraints were already summarized
in the previous section.

An interesting outcome of this scrutiny is the possibility, 
pointed out by us previously~\cite{Fuyuto:2014cya}, that
an invisible $Z'$ boson could evade the commonly accepted 
Grossman-Nir (GN) bound~\cite{Grossman:1997sk} of 
${\cal B} (K_L\to\pi^0\nu\bar\nu) \lesssim 1.4 \times 10^{-9}$.

\subsection{$t\to cZ'(\to \nu\bar\nu)$ via left-handed current}

We use BaBar data on $B\to K^{(*)}\nu\bar\nu$ in Eq. (\ref{eq:B2KZp-BaBar})
to constrain the tree-level effective $bsZ'$ couplings $g_{sb}^L$ and $g_{sb}^R$.
The $2\sigma$ range for $B^+\to K^+\nu\bar\nu$ imposes
\begin{align}
0.16\times 10^{-9}\lesssim |g_{sb}^{L}+g_{sb}^{R}|
\left(\frac{100\,\MeV}{m_{Z'}}\right) \lesssim 0.88\times 10^{-9},
\end{align}
while the $B^+\to K^{*+}\nu\bar\nu$ data mainly constrains the other combination of
the $bsZ'$ couplings,
\begin{align}
|g_{sb}^{L}-g_{sb}^{R}|\left(\frac{100~\MeV}{m_{Z'}}\right) \lesssim 1.3\times 10^{-9}.
\end{align}
Combining the two constraints, we get
\begin{align}
|g_{sb}^{L}|, |g_{sb}^{R}| \lesssim 1.1\times 10^{-9}\left(\frac{m_{Z'}}{100~\MeV}\right),
\end{align}
for $m_{Z'}<2m_\mu$.
Note that the excess in the $B^+\to K^+\nu\bar\nu$ data does not necessarily imply
a nonzero $g_{sb}^L$, as $g_{sb}^L=0$ can still explain the excess by a nonzero $g_{sb}^R$.

Using the SU(2)$_L$ relation $g_{ct}^L\simeq g_{sb}^L$,
we obtain the upper limit on the left-handed current 
contribution to the $t\to cZ'$ branching ratio,
\begin{align}
{\cal B}(t\to cZ')_{\rm LH} \lesssim 4\times 10^{-12}
\end{align}
for $m_{Z'} <2m_\mu$.
Note that the limit does not depend on $m_{Z'}$, as it cancels out in the final expression.

%
\begin{figure}[t!]
{
 \includegraphics[width=75mm]{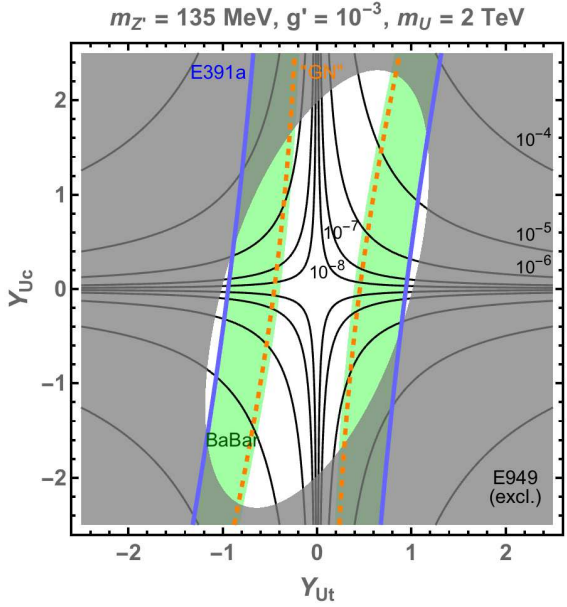}
}
\caption{
Same as Fig. \ref{fig:219-334_Y} [right], but for $m_{Z^\prime}=135$ MeV and $g'=10^{-3}$,
with the E949 exclusion of Eq.~(\ref{eq:E949-pi0}).
} \label{fig:135_Y}
\end{figure}
\subsection{$t\to cZ'(\to \nu\bar\nu)$ via right-handed current}

We constrain the right-handed current contribution to the $t\to cZ'$ branching ratio
by using $B^+\to K^+\nu\bar\nu$ and $K\to\pi\nu\bar\nu$ data.
As discussed in the previous section, the E949 constraint from
$K^+\to \pi^+P^0(\to \nu\bar\nu)$ search can be avoided, if the $Z'$ mass falls into
the $\pi^0$-mass window, i.e. 125 MeV $\lesssim m_{Z'}\lesssim$ 150 MeV.
Although this mass window is still constrained by $\pi^0\to\nu\bar\nu$,
searched in $K^+\to\pi^+\pi^0$ [Eq.~(\ref{eq:E949-pi0})], the limit is rather weak
compared to the $K^+\to \pi^+P^0(\to \nu\bar\nu)$ limits outside the $\pi^0$-window.
[See explanation below Eq.~(\ref{eq:E949-260}).]
To allow for the possibility of a large $t\to cZ'$ rate,
we take the following benchmark point:
\begin{itemize}
\item $m_{Z'}=135$ MeV, $g'=10^{-3}$, $m_U=2$ TeV,
\end{itemize}
which is shown by a red cross in Fig.~\ref{fig:lifetime}.
The $B$ and $K$ decay constraints are summarized in Table~\ref{tab:exp}.
As discussed in the previous section, this particular choice of $g'$ and $m_U$ does not
affect the final result for ${\cal B}(t\to cZ')_{\rm RH}$.

In Fig.~\ref{fig:135_Y}, we give contours of ${\cal B}(t\to cZ')_{\rm RH}$ as black-solid
lines in the $(Y_{Ut},Y_{Uc})$ plane.
The $B$ and $K$ decay constraints are overlaid with the shadings and line styles
as in Fig.~\ref{fig:219-334_Y} [right].
The semi-transparent dark-gray shaded region is excluded by the E949 limit on
$\pi^0\to \nu\bar\nu$ at 90\% C.L. [Eq.~(\ref{eq:E949-pi0})].
In the present case, the E391a constraint on $K_L\to\pi^0\nu\bar\nu$
 [Eq.~(\ref{eq:E391a})], shown as blue-solid lines, also plays a role.
The green shaded regions, favored by the mild $B^+\to K^+\nu\bar\nu$ excess
in BaBar data [in Eq.~(\ref{eq:B2KZp-BaBar})], are compatible with other constraints
in most parts of the shown range. 
This is in contrast to Scenario (ii-a), where the constraints from 
$B^0\to K^{*0}\chi(\to \mu^+\mu^-)$ and
$K^+\to \pi^+ P^0(\to \nu\bar\nu)$ exclude the BaBar regions.
In this benchmark point, we obtain ${\cal B}(t\to cZ')_{\rm RH}\lesssim 5\times 10^{-5}$.

\begin{figure}[t!]
{\includegraphics[width=75mm]{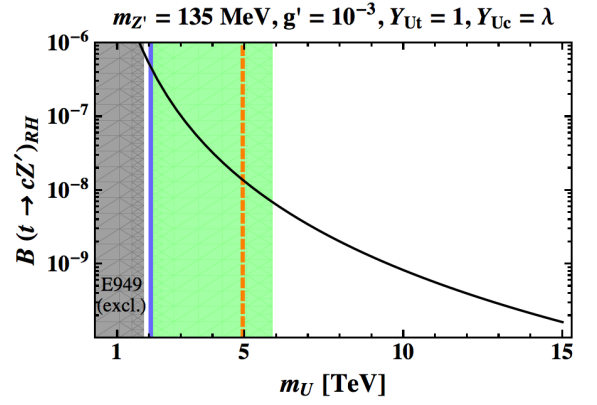} 
}
\vskip0.05cm
\caption{
Same as Fig. \ref{fig:219-334_mU} [right], but for $m_{Z^\prime}=135$ MeV and $g'=10^{-3}$.
} \label{fig:135_mU}
\end{figure}

Fixing the Yukawa couplings to $Y_{Ut}=1$, $Y_{Uc}=\lambda$,
we plot ${\cal B}(t\to cZ')_{\rm RH}$ in Fig.~\ref{fig:135_mU} 
as a function of $m_U$ with the same $m_{Z'}$ and $g'$ values.
For this case, the BaBar excess favors a nonzero but small $t\to cZ'$ rate within
$6\times 10^{-9} \lesssim {\cal B}(t\to cZ')_{\rm RH}\lesssim 5\times 10^{-7}$.

For the $Z'$ mass within the $\pi^0$-window, i.e. 125 MeV $\lesssim m_{Z'}\lesssim$
150 MeV, the same E949 limit for $K^+\to\pi^+Z'$ applies and we obtain similar results
for ${\cal B}(t\to cZ')_{\rm RH}$.

The E949 limit gets stronger considerably if the $Z'$ mass is out of the $\pi^0$-window,
hence ${\cal B}(t\to cZ')_{\rm RH}$ cannot be larger than the above limits.
For instance, taking $m_{Z'}=11$ MeV with $g'=5\times 10^{-4}$, motivated by
IceCube data~\cite{Araki:2014ona,Araki:2015mya}, we obtain
${\cal B}(t\to cZ')_{\rm RH}\lesssim 6\times 10^{-8}$.

In summary, we obtain ${\cal B}(t\to cZ')_{\rm RH}\lesssim 5\times 10^{-5}$
for any $m_{Z'}$ in Scenario (ii-b).
The decay $t\to cZ'(\to \nu\bar\nu\ {\rm at\ 100\%})$ with such a small branching ratio
might be quite challenging for searches at the LHC.

%
\subsection{Apparent violation of Grossman-Nir bound}

The light $Z'$ has an interesting implication for 
kaon decay experiments~\cite{Fuyuto:2014cya}.

From isospin symmetry, the branching ratio ${\cal B}(K_L\to\pi^0\nu\bar\nu)$
is connected with ${\cal B}(K^+\to\pi^+\nu\bar\nu)$ by 
a {\it model-independent} relation, known as the Grossman-Nir (GN) bound~\cite{Grossman:1997sk},
\begin{align}
\mathcal B (K_L\to\pi^0\nu\bar\nu)
\lesssim  4.3\times \mathcal B (K^+\to\pi^+\nu\bar\nu), \label{eq:exact-GN}
\end{align}
where the overall factor of 4.3 comes from $\tau_{K_L}/\tau_{K^+}\simeq 4.1$
and isospin-breaking effects.
Plugging in the 90\% C.L. upper limit of ${\cal B}(K^+\to \pi^+\nu\bar\nu)<3.35 \times
10^{-10}$ by E949~\cite{Artamonov:2009sz},
the GN bound leads to
\begin{align}
{\cal B} (K_L\to\pi^0\nu\bar\nu) \lesssim 1.4 \times 10^{-9}.\ ({\rm ``GN"})
 \label{eq:usual-GN}
\end{align}
This is an order of magnitude stronger than the direct 
limit on $K_L\to\pi^0\nu\bar\nu$ by E391a, Eq.~(\ref{eq:E391a}).
In Figs.~\ref{fig:219-334_Y}--\ref{fig:135_mU}, this commonly accepted GN bound is
shown by the orange-dashed lines.

There are two ongoing experiments in search for $K\to\pi\nu\bar\nu$ decays.
The NA62 experiment~\cite{NA62} at CERN aims at
measuring of order 100 $K^+\to \pi^+\nu\bar\nu$ events,
while the KOTO experiment \cite{KOTO} at J-PARC aims at 
3$\sigma$ measurement of $K_L\to\pi^0\nu\bar\nu$ at SM rate.
KOTO has already reached~\cite{KOTO-CKM2014} the sensitivity of
E391a [Eq.~(\ref{eq:E391a})], 
but folklore is that KOTO can start to probe New Physics effects 
only after Eq.~(\ref{eq:usual-GN}) is breached.

We have argued, however, that the kinematic selection in $K^+\to \pi^+\nu\bar\nu$
searches (including both E949 and NA62) makes them insensitive to 
the possible existence of a light new boson $X^0$, 
produced in $K^+\to\pi^+X^0$, if $m_{X^0} \sim m_{\pi}$ or larger than $2m_\pi$.
If so, the usual GN bound of Eq.~(\ref{eq:usual-GN}) does not apply,
and therefore, without such a selection, KOTO is already entering
the domain of New Physics.

%
\begin{figure}[t!]
{\includegraphics[width=75mm]{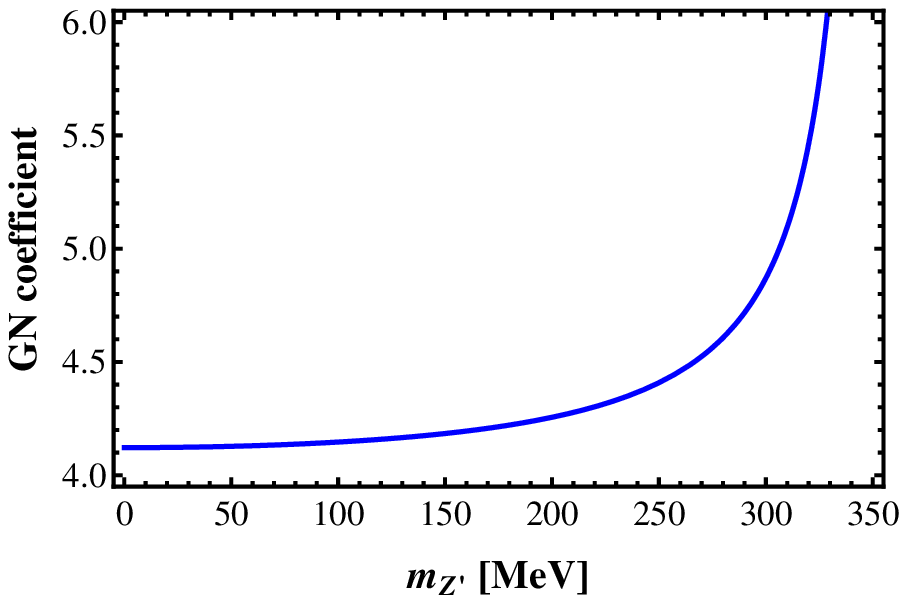} 
}
\caption{
Maximally allowed value as a function of $m_{Z'}$ for
ratio ${\cal B}(K_L\to \pi^0Z')/{\cal B}(K^+\to \pi^+Z')$,
given in Eq. (\ref{eq:GN-Zp}).
} \label{fig:GN-Zp}
\end{figure}

A relation similar to Eq.~(\ref{eq:exact-GN}) still holds for the light $Z'$ contribution,
with a slight modification in the overall coefficient.
Taking the ratio of the $K_L\to \pi^0Z'$ and $K^+\to \pi^+Z'$ branching ratios
in Eq.~(\ref{eq:BR-K2piZp}), we obtain the light $Z'$ version of the GN bound:
\begin{align}
\frac{\mathcal B(K_L\to \pi^0 Z')}{\mathcal B(K^+\to \pi^+ Z')}
&=\frac{\tau_{K_L}}{\tau_{K^+}} \frac{1}{r(m_{Z'}^2)}
 \left|\frac{{\rm Im}(g_{ds}^L+g_{ds}^R)}{g_{ds}^L+g_{ds}^R} \right|^2 \notag\\
&\leq \frac{\tau_{K_L}}{\tau_{K^+}} \frac{1}{r(m_{Z'}^2)}, \label{eq:GN-Zp}
\end{align}
where the isospin breaking factor $r(m_{Z'}^2)$ is defined by
\begin{align}
\frac{1}{r(m_{Z^\prime}^2)}
&\equiv \frac{m_{K_L}^3}{m_{K^+}^3}
 \frac{\beta_{K_L\pi^0 Z'}^3}{\beta_{K^+\pi^+ Z'}^3}
 \left[\frac{f_+^{K^0\pi^0}(m_{Z'}^2)}{f_+^{K^+\pi^+}(m_{Z'}^2)}\right]^2.
\end{align}
To keep generality, we recover the dependence on $g_{ds}^R$, assuming
the interaction form of Eq.~(\ref{eq:Leff-Z'}).
The form factor ratio is known to be $q^2$-independent at next-to-leading order
in chiral perturbation theory, with numerical value
$f_+^{K^+\pi^+}(q^2)/ f_+^{K^0\pi^0}(q^2) =1.0238\pm 0.0022$ \cite{Mescia:2007kn}.
The genuine GN bound of Eq.~(\ref{eq:GN-Zp}) is then given by
\begin{align}
\frac{ {\cal B}(K_L\to \pi^0Z')}{ {\cal B}(K^+\to \pi^+Z')}
&\lesssim 4.122\left( \frac{1.003}{r(m_{Z'}^2)} \right),
\end{align}
where $r(0)=1.003$ is taken as reference.
The right-hand side, the ``GN coefficient", 
depends on the $Z'$ mass, as illustrated in Fig.~\ref{fig:GN-Zp}.

For $125~\MeV \lesssim m_{Z'}\lesssim 150~\MeV$, plugging in the 90\% C.L. upper limit
on ${\cal B}(K^+\to \pi^+Z')$ by E949 [Eq. (\ref{eq:E949-pi0})], we obtain
${\cal B}(K_L\to \pi^0Z') \lesssim 2.3\times 10^{-7}$.
The direct bound on $K_L\to\pi^0\nu\bar\nu$ by E391a, Eq. (\ref{eq:E391a}),
is indeed stronger than this true GN bound.

The above argument is general and applicable to any weakly interacting
light boson, or short-lived invisibly decaying boson, that couples to $s\to d$ currents.
The bound of Eq.~(\ref{eq:GN-Zp}) holds for a massive vector boson
that couples to the $s\to d$ currents in the form of Eq.~(\ref{eq:Leff-Z'}).
On the other hand, for the $L_\mu-L_\tau$ gauge boson with loop-induced $sdZ'$
coupling of Eq.~(\ref{eq:eff-coup-loop}), we obtain
$|{\rm Im}(\Delta g_{ds}^L)/g_{ds}^L|^2 \sim |{\rm Im}(V_{ts}V_{td}^*)/(V_{ts}V_{td}^*)|^2
\simeq 0.15$, as long as $Y_{Ut}\gtrsim Y_{Uc}$, due to top-top dominance in loop.
Thus, the GN bound of Eq.~(\ref{eq:GN-Zp}) cannot be saturated in this case.

The argument can be further extended to three-body kaon decays where the final state
contains a pair of new massive invisible particles ($\chi$), i.e., $K\to\pi\chi\chi$.
If the mass of $\chi$ is larger than $m_\pi$, the decay is allowed only
if the invariant mass of the $\chi$ pair satisfies $2m_\pi<m_{\chi\chi}(<m_K-m_\pi)$.
In this case, the $\pi^+$ momentum is always outside the signal regions of
the $K^+\to\pi^+\nu\bar\nu$ experiments, hence, the usual GN bound does not apply.
An interesting candidate is the very light neutralino in the minimal supersymmetric
standard model, which was discussed in Ref.~\cite{Dreiner:2009er}, 
although the analysis needs to be updated in light of recent LHC results.
(See Ref.~\cite{deVries:2015mfw} for recent assessment of the light neutralino.)

\section{Discussion and Conclusions \label{sec:disc}}

The so-called $P_5^\prime$ and $R_K$ anomalies in $b\to s$ transitions, 
as revealed by LHCb data, suggest the possible existence of
a new massive gauge boson $Z'$ coupling to left-handed $b\to s$ current, 
which in turn implies $tcZ'$ coupling.
Motivated by this, we studied the top FCNC decay $t\to cZ'$ 
based on the gauged $L_\mu -L_\tau$ model with 
vector-like quarks that mix with SM quarks.
The model can also be applied to address the muon $g-2$ anomaly,
which turns out to allow only a very light $Z'$ due to neutrino scattering data.
The situation is mutually exclusive with the $b\to s$ anomalies.  
We studied how large the $t\to cZ'$ rate can be in three well-motivated scenarios:
(i)~heavy $Z'$ with $m_b\lesssim m_{Z'}<m_t-m_c$, 
  motivated by the $P_5'$ and $R_K$ anomalies;
(ii-a) light $Z'$ with $2m_\mu < m_{Z'}\lesssim 400$ MeV,
  motivated by the muon $g-2$ anomaly; 
(ii-b) the $(g-2)_\mu$-motivated $Z'$ with $m_{Z'}<2m_\mu$.

In Scenario (i), using a global fit result of $b\to s$ data 
as well as $B_s$ meson mixing constraint, we find that 
the left-handed current contribution to branching ratio 
${\cal B}(t\to cZ')_{\rm LH}$ can be as large as $10^{-6}$.
We also find that the right-handed current contribution ${\cal B}(t\to cZ')_{\rm RH}$, 
which is not constrained by $B$ data, can be as large as ${\cal O}(10^{-4})$ 
with reasonably large mixing (around the Cabibbo angle) 
between vector-like quark $U$ and $t$, $c$.
The left-handed case would be beyond the reach of even the high-luminosity LHC upgrade,
while the right-handed case might be accessible already with LHC Run 1 data.
[See Eq.~(\ref{eq:t2cZp-CMS}) for our naive projection 
based on $t\to qZ$ results.]

In Scenario (ii-a), we find ${\cal B}(t\to cZ')_{\rm LH}$ to be extremely tiny,
below $10^{-11}$, due to rare $B$ decay constraints.
In this scenario, even the right-handed current contribution is constrained by rare $B$ and 
$K$ decays via one-loop effects.
We find that ${\cal B}(t\to cZ')_{\rm RH}\gtrsim 10^{-6}$ is 
allowed only at the cost of fine-tuning the relation between $Y_{Ut}$ and $Y_{Uc}$.
Nevertheless, in such cancellation regions, ${\cal B}(t\to cZ')_{\rm RH}$ may 
be larger than ${\cal O}(10^{-5})$ for $330~\MeV \lesssim m_{Z'}\lesssim 400$ MeV.
Our naive projection based on $t\to qZ$ results suggests that this could be within
reach of the ATLAS and CMS experiments with 300$^{-1}$ data at the (13-)14 TeV LHC.
However, a careful collider study is needed 
to find the true sensitivity, as the search strategy 
needs to be changed from the $t\to qZ$ case.

Scenario (ii-b) can accommodate larger $t\to cZ'$ branching ratios for the
right-handed current contribution: ${\cal B}(t\to cZ')_{\rm RH}\lesssim 5\times 10^{-5}$.
This case, however, would be more challenging for collider search, as
the $Z'$ decays exclusively into neutrinos (but with little missing mass).
Such a light $Z'$ is interesting instead for rare kaon decay experiments,
and could even lead to observation of New Physics beyond the
so-called Grossman-Nir bound, or ${\cal B}(K_L \to \pi^0 + {\rm nothing}) > 1.4 \times 10^{-9}$.
If this happens, our prediction is that it occurs via
$K_L \to \pi^0 X^0$ with unobserved $m_{X^0} \sim m_{\pi^0}$,
with our $Z'$ motivated by muon $g-2$ as a candidate.
We remark that the $Z'$ in Scenarios (ii-a) and (ii-b) may also be probed by
the future neutrino beam facility LBNE~\cite{LBNE}
via neutrino trident production~\cite{Altmannshofer:2014pba}.
And certainly LHCb and Belle II experiments should pursue further
``bump" searches in $B \to K^{(*)}\mu^+\mu^-$ and $B \to K^{(*)}\nu\bar\nu$ decays.

In this paper, we assumed a particular $Z'$ model to study $t\to cZ'$ decay.
In the model, the right-handed $tcZ'$ coupling correlates with the $ttZ'$ and $ccZ'$couplings.
In particular, the $ttZ'$ coupling is strongly constrained by loop-induced decays
from chiral $m_t/m_c$ enhancement, and the $tcZ'$ coupling in turn is also constrained
indirectly.
This correlation is not general, and the meson decay constraints might be relaxed
in some other $Z'$ models where the right-handed $tcZ'$ coupling is independent from
the $ttZ'$ coupling.


The muon $g-2$ anomaly implies that the U(1)$'$ symmetry breaking scale $v_\Phi$
is around the electroweak scale of $246$ GeV or below.
The mass of the new Higgs boson $\phi$ behind the 
spontaneous breaking of the U(1)$'$ symmetry, 
is hence expected to be below 1 TeV and within reach at the LHC.
The mixing of the vector-like $U$ quark with top via the $\phi$-Yukawa interaction
leads to effective $tt\phi$ coupling.
Thus, the $\phi$ can be produced via gluon fusion $gg\to \phi$, followed by
$\phi\to Z'Z'(\to 4\mu/2\mu2\nu)$, as pointed out in Ref.~\cite{Fuyuto:2014cya}.
The effective $tt\phi$ coupling, however, is highly suppressed compared to
the SM top Yukawa coupling, due to the constraints from
$B^0\to K^{*0}Z'(\to \mu^+\mu^-)$ as discussed in Sec.~\ref{sec:heavy},
hence the $gg\to \phi$ cross section is too small to be observed at the LHC~\cite{HK:phi}.
Instead, the effective $tc\phi$ couplings, generated in a similar way as $tcZ'$,
may offer another $\phi$ production mechanism, i.e. $t\to c\phi$ in
$t\bar t$ events at the LHC. 
This gives rise to a striking signature, namely two collimated dimuons in
$pp\to t\bar t\to bWc\phi(\to Z'Z')$ with $Z'Z'\to(\mu^+\mu^-)(\mu^+\mu^-)$.
This interesting possibility will be pursued elsewhere~\cite{HK:phi}.

\vskip0.3cm
\noindent{\bf Acknowledgement}.
KF is supported by Research Fellowships of the Japan Society 
for the Promotion of Science for Young Scientists, No. 15J01079.
WSH is supported by the Academic Summit grant MOST 103-2745-M-002-001-ASP of the
Ministry of Science and Technology, as well as by grant NTU-EPR-103R8915.
MK is supported under NSC 102-2112-M-033-007-MY3.
MK thanks Y.~Chao, A.~Mauri, J.~Tandean and M.~Williams for valuable discussions.
WSH thanks T. Blake and M. Pepe-Altarelli for correspondence.
KF thanks useful correspondence with S. Gori, and
the NTUHEP group for hospitality during exchange visits.

\appendix

\section{ Efficiency for $B^0\to K^{*0}Z' \to K\pi\mu^+\mu^-$ 
\label{app:B2Kpimumu}}

In order to estimate the efficiency for the $B^0\to K^{*0}Z' \to K\pi\mu^+\mu^-$ decay
at the LHCb, we need information of the angular distribution for this decay.
In the narrow width approximation, the normalized 
differential decay width for $\bar B^0\to \bar K^{*0}Z'\to K^-\pi^+\mu^+\mu^-$
is given by
\begin{align}
&\frac{1}{\Gamma}
 \frac{d\Gamma}{d\cos\theta_K d\cos\theta_\ell d\phi} \notag\\
&= \frac{9}{16\pi (1+2m_\mu^2/m_{Z'}^2) (|H_0|^2+|H_+|^2+|H_-|^2)} \notag\\
 &\quad \times \bigg\{ -\beta_\mu^2 \bigg[ |H_0|^2\cos^2\theta_K\cos^2\theta_\ell
 \notag\\
 &\quad  +\frac{1}{4} (|H_+|^2 +|H_-|^2) \sin^2\theta_K\sin^2\theta_\ell
 +\Xi(\theta_K,\theta_\ell, \phi) \bigg] \notag\\
 &\quad + |H_0|^2\cos^2\theta_K
 +\frac{1}{2} (|H_+|^2 +|H_-|^2) \sin^2\theta_K \bigg\}, \label{eq:B2Kpimumu}
\end{align}
where $\beta_\mu \equiv\sqrt{1-4m_\mu^2/m_{Z'}^2 }$ and
the helicity amplitudes $H_{0,\pm}$ are given in Eq. (\ref{eq:Hm}).
The $\phi$-dependence enters solely through the function
\begin{align}
&\Xi(\theta_K,\theta_\ell, \phi) \notag\\
&= -\frac{1}{4}\sin 2\theta_K\sin 2\theta_\ell \big\{
 \cos\phi \left[ {\rm Re}(H_0H_+^*) +{\rm Re}(H_0H_-^*) \right] \notag\\
 &\quad -\sin\phi \left[ {\rm Im}(H_0H_+^*) -{\rm Im}(H_0H_-^*) \right] \big\}
 +\frac{1}{2}\sin^2\theta_K\sin^2\theta_\ell \notag\\
 &\quad \times \left[ \cos 2\phi {\rm Re}(H_+H_-^*) +\sin 2\phi {\rm Im}(H_+H_-^*) \right].
\end{align}
We follow the LHCb convention~\cite{Aaij:2013iag} for the definition of decay angles:
$\theta_\ell$ is the angle between direction of $\mu^-$ and direction opposite
to $\bar B^0$ in the $Z'$ rest frame;
$\theta_K$ is the angle between direction of $K^-$ and
direction opposite to $\bar B^0$ in the $\bar K^{*0}$ rest frame;
$\phi$ is the angle between $Z'\to \mu^+\mu^-$ decay plane and
$K^{*0}\to K^-\pi^+$ decay plane in the $\bar B^0$ rest frame.

\begin{figure}[t!]
{
 \includegraphics[width=75mm]{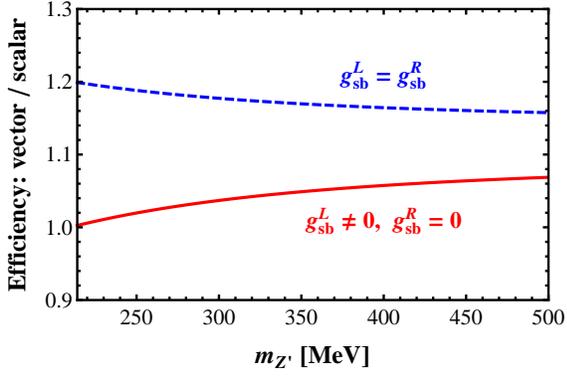}
}
\caption{
Ratio of efficiencies between vector-boson $Z'$ and scalar-boson $\chi$ 
for $\bar B^0\to \bar K^{*0}Z'(\chi)\to K^-\pi^+\mu^+\mu^-$ events collected by LHCb.
Solid line is for $g_{sb}^L\neq 0$, $g_{sb}^R=0$ and
dashed line is for $g_{sb}^L =g_{sb}^R$.
The other two cases of $g_{sb}^L= 0$, $g_{sb}^R\neq 0$ and $g_{sb}^L= -g_{sb}^R$
behave similarly to the solid line.
} \label{fig:eff}
\end{figure}

If a new scalar-boson $\chi$ mediates the four-body decay instead of the vector-boson
$Z'$, the angular distribution simply behaves as
$d\Gamma/d\cos\theta_K d\cos\theta_\ell d\phi \propto \cos^2\theta_K$.
This is the case assumed in the LHCb search~\cite{Aaij:2015tna} for 
hidden-sector bosons $\chi$ in $B^0\to K^{*0}\chi$.
In order to convert the limits on $B^0\to K^{*0}\chi$ into the $Z'$ case,
one needs to know the ratio of the efficiencies between the $\chi$ and $Z'$ cases.
Ref.~\cite{Aaij:2015tna} (see Supplemental Material) provides this information
in the form of ratio between integrals of the trigonometric functions appearing in 
Eq.~(\ref{eq:B2Kpimumu}) and integral of $\cos^2\theta_K$, taking into account the efficiency.
Using this information, we obtain the ratio of efficiencies for $Z'$ to $\chi$,
as shown in Fig.~\ref{fig:eff}.
For $g_{sb}^R=0$, corresponding to the loop-induced coupling discussed in
Sec.~\ref{sec:light-mumu}, the change in efficiencies from the scalar case
is within 6\% for $m_{Z'}\leq 400$ MeV.
If we allow general chiral structure for $bsZ'$ coupling, the change is still small,
within 20\% for $m_{Z'}\leq 400$ MeV.

\

\section{Loop functions for effective 
 couplings \label{app:loop}}

The loop functions given in Eq.~(\ref{eq:loop-func-apprx}) are 
approximate formulas in the large $m_U$ limit.
In our numerical study, we use the following expression:
\begin{align}
f_{qq'} &= -4m_W^2 m_U^4 I_0^{qq'} +(2m_W^2 +m_U^2)m_U^2 I_2^{qq'} -2m_U^2 I_4^{qq'},
\label{eq:loop-exact}
\end{align}
where $q,q'=t,c$, and
\begin{align}
&I_0^{qq'} \notag\\
&\equiv \int\frac{d^4k}{i(2\pi)^4} \frac{16\pi^2}
 {(k^2-m_q^2)(k^2-m_{q'}^2)(k^2-m_{U}^2)^2(k^2-m_W^2)} \notag\\
&= - \int_0^1 dx_1 \int_0^{1-x_1}dx_2~\frac{(1-x_1-x_2)^2}{\alpha_{qq'}\beta_{qq'}^2}, 
\end{align}
\begin{align}
&I_2^{qq'} \notag\\
&\equiv \int\frac{d^4k}{i(2\pi)^4} \frac{16\pi^2 k^2}
 {(k^2-m_q^2)(k^2-m_{q'}^2)(k^2-m_{U}^2)^2(k^2-m_W^2)} \notag\\
&= \frac{2}{(m_{U}^2-m_W^2)^2}\int_0^1 dx_1 \int_0^{1-x_1}dx_2 \notag\\
 &\quad \times \left[ \ln\frac{\beta_{qq'}}{\alpha_{qq'}}
  -\frac{(1-x_1-x_2)(m_{U}^2-m_W^2)}{\beta_{qq'}}\right], 
\end{align}
\begin{align}
&I_4^{qq'} \notag\\
&\equiv \int\frac{d^4k}{i(2\pi)^4} \frac{16\pi^2(k^2)^2}
 {(k^2-m_q^2)(k^2-m_{q'}^2)(k^2-m_{U}^2)^2(k^2-m_W^2)} \notag\\
&=-\frac{6}{m_{U}^2-m_W^2}\int_0^1 dx_1 \int_0^{1-x_1}dx_2 \notag\\
 &\quad \times \left(
 1-x_1-x_2 -\frac{\alpha_{qq'}}{m_{U}^2-m_W^2}\ln\frac{\beta_{qq'}}{\alpha_{qq'}} \right),
\end{align}
with
\begin{align}
&\alpha_{qq'}=x_1m_q^2 +x_2m_{q'}^2 +(1-x_1-x_2)m_W^2, \notag\\
&\beta_{qq'} =x_1m_q^2 +x_2m_{q'}^2 +(1-x_1-x_2)m_{U}^2.
\end{align}




\begin{thebibliography}{99}

\bibitem{Agashe:2014kda}
  K.A.~Olive {\it et al.}  [Particle Data Group],
  Chin.\ Phys.\ C {\bf 38}, 090001 (2014).


\bibitem{Aaij:2013qta}
  R.~Aaij {\it et al.}  [LHCb Collaboration],
  Phys.\ Rev.\ Lett.\  {\bf 111}, 191801 (2013)
  [arXiv:1308.1707 [hep-ex]].

\bibitem{LHCb:2015dla}
  LHCb Collaboration,
  LHCb-CONF-2015-002; 
   R.~Aaij {\it et al.} [LHCb Collaboration],
  JHEP {\bf 1602}, 104 (2016)
  [arXiv:1512.04442 [hep-ex]].

\bibitem{Aaij:2014ora}
  R.~Aaij {\it et al.}  [LHCb Collaboration],
  Phys.\ Rev.\ Lett.\  {\bf 113}, 151601 (2014)
  [arXiv:1406.6482 [hep-ex]].


\bibitem{Descotes-Genon:2013wba}
  S.~Descotes-Genon, J.~Matias, J.~Virto,
  Phys.\ Rev.\ D {\bf 88}, 074002 (2013)
  [arXiv:1307.5683 [hep-ph]].

\bibitem{Altmannshofer:2013foa}
  W.~Altmannshofer and D.M.~Straub,
  Eur.\ Phys.\ J.\ C {\bf 73}, 2646 (2013)
  [arXiv:1308.1501 [hep-ph]].

\bibitem{Beaujean:2013soa}
  F.~Beaujean, C.~Bobeth, D.~van Dyk,
  Eur.\ Phys.\ J.\ C {\bf 74}, 2897 (2014)
  [Erratum Eur.\ Phys.\ J.\ C {\bf 74}, 3179 (2014)]
  [arXiv:1310.2478 [hep-ph]].

\bibitem{Horgan:2013pva}
  R.R.~Horgan, Z.~Liu, S.~Meinel, M.~Wingate,
  Phys.\ Rev.\ Lett.\  {\bf 112}, 212003 (2014)
  [arXiv:1310.3887 [hep-ph]].

\bibitem{Hurth:2013ssa}
  T.~Hurth and F.~Mahmoudi,
  JHEP {\bf 1404}, 097 (2014)
  [arXiv:1312.5267 [hep-ph]].

\bibitem{Alonso:2014csa}
  R.~Alonso, B.~Grinstein, J.M. Camalich,
  Phys.\ Rev.\ Lett.\  {\bf 113}, 241802 (2014)
  [arXiv:1407.7044 [hep-ph]].

\bibitem{Hiller:2014yaa}
  G.~Hiller and M.~Schmaltz,
  Phys.\ Rev.\ D {\bf 90}, 054014 (2014)
  [arXiv:1408.1627 [hep-ph]].

\bibitem{Ghosh:2014awa}
  D.~Ghosh, M.~Nardecchia, S.A.~Renner,
  JHEP {\bf 1412}, 131 (2014)
  [arXiv:1408.4097 [hep-ph]].

\bibitem{Hurth:2014vma}
  T.~Hurth, F.~Mahmoudi, S.~Neshatpour,
  JHEP {\bf 1412}, 053 (2014)
  [arXiv:1410.4545 [hep-ph]].

\bibitem{Altmannshofer:2014rta}
  W.~Altmannshofer and D.M.~Straub,
  Eur.\ Phys.\ J.\ C {\bf 75}, 382 (2015)
  [arXiv:1411.3161 [hep-ph]].


\bibitem{Altmannshofer:2014cfa}
  W.~Altmannshofer, S.~Gori, M.~Pospelov, I.~Yavin,
  Phys.\ Rev.\ D {\bf 89}, 095033 (2014)
  [arXiv:1403.1269 [hep-ph]].

\bibitem{He:1990pn}
  X.-G.~He, G.C.~Joshi, H.~Lew, R.R.~Volkas,
  Phys.\ Rev.\ D {\bf 43}, 22 (1991).

\bibitem{Fox:2011qd}
  P.J.~Fox, J.~Liu, D.~Tucker-Smith, N.~Weiner,
  Phys.\ Rev.\ D {\bf 84}, 115006 (2011)
  [arXiv:1104.4127 [hep-ph]].

\bibitem{Jung:2009jz}
  The $t\to uZ'$ decay with a hadronically decaying $Z'$ was discussed
  in S.~Jung, H.~Murayama, A.~Pierce, J.D.~Wells,
  Phys.\ Rev.\ D {\bf 81}, 015004 (2010)
  [arXiv:0907.4112 [hep-ph]].


\bibitem{Jegerlehner:2009ry}
  F.~Jegerlehner and A.~Nyffeler,
  Phys.\ Rept.\  {\bf 477}, 1 (2009)
  [arXiv:0902.3360 [hep-ph]].

\bibitem{Baek:2001kca}
  S.~Baek, N.G.~Deshpande, X.-G.~He, P.~Ko,
  Phys.\ Rev.\ D {\bf 64}, 055006 (2001)
  [hep-ph/0104141].

\bibitem{Altmannshofer:2014pba}
  W.~Altmannshofer, S.~Gori, M.~Pospelov, I.~Yavin,
  Phys.\ Rev.\ Lett.\  {\bf 113}, 091801 (2014)
  [arXiv:1406.2332 [hep-ph]].


\bibitem{Fuyuto:2014cya}
  K.~Fuyuto, W.-S.~Hou, M.~Kohda,
  Phys.\ Rev.\ Lett.\  {\bf 114}, 171802 (2015)
  [arXiv:1412.4397 [hep-ph]].

\bibitem{Grossman:1997sk}
  Y.~Grossman and Y.~Nir,
  Phys.\ Lett.\ B {\bf 398}, 163 (1997)
  [hep-ph/9701313].


\bibitem{Araki:2014ona}
  T.~Araki, F.~Kaneko, Y.~Konishi, T.~Ota, J.~Sato, T.~Shimomura,
  Phys.\ Rev.\ D {\bf 91}, 037301 (2015)
  [arXiv:1409.4180 [hep-ph]].

\bibitem{Aartsen:2014gkd}
  M.G.~Aartsen {\it et al.} [IceCube Collaboration],
  Phys.\ Rev.\ Lett.\  {\bf 113}, 101101 (2014)
  [arXiv:1405.5303 [astro-ph.HE]].


\bibitem{HK:phi}
  W.-S.~Hou and M.~Kohda, work in progress.



\bibitem{Mishra:1991bv}
  S.R.~Mishra {\it et al.}  [CCFR Collaboration],
  Phys.\ Rev.\ Lett.\  {\bf 66}, 3117 (1991).


\bibitem{Aad:2014wra}
  G.~Aad {\it et al.}  [ATLAS Collaboration],
  Phys.\ Rev.\ Lett.\  {\bf 112}, 231806 (2014)
  [arXiv:1403.5657 [hep-ex]].

\bibitem{Harigaya:2013twa}
  For a recent study of $Z'$ search in $Z\to 4\ell$, see also
  K.~Harigaya, T.~Igari, M.M.~Nojiri, M.~Takeuchi, K.~Tobe,
  JHEP {\bf 1403}, 105 (2014)
  [arXiv:1311.0870 [hep-ph]].


\bibitem{Crivellin:2015mga}
  A.~Crivellin, G.~D'Ambrosio, J.~Heeck,
  Phys.\ Rev.\ Lett.\  {\bf 114}, 151801 (2015)
  [arXiv:1501.00993 [hep-ph]].


\bibitem{Aad:2015uza} 
  G.~Aad {\it et al.} [ATLAS Collaboration],
  Eur.\ Phys.\ J.\ C {\bf 76}, 12 (2016)
  [arXiv:1508.05796 [hep-ex]].

\bibitem{Chatrchyan:2013nwa}
  S.~Chatrchyan {\it et al.}  [CMS Collaboration],
  Phys.\ Rev.\ Lett.\  {\bf 112}, 171802 (2014)
  [arXiv:1312.4194 [hep-ex]].

\bibitem{CMS:2013zfa}
  CMS Collaboration, 
  CMS-PAS-FTR-13-016.

\bibitem{CMS:2013xfa}
  CMS Collaboration,
  arXiv:1307.7135 [hep-ex].

\bibitem{Chatrchyan:2012hqa}
  S.~Chatrchyan {\it et al.} [CMS Collaboration],
  Phys.\ Lett.\ B {\bf 718}, 1252 (2013)
  [arXiv:1208.0957 [hep-ex]].

\bibitem{ATLAS:2013hta}
  ATLAS Collaboration,
  arXiv:1307.7292 [hep-ex].

\bibitem{Agashe:2013hma}
  K.~Agashe {\it et al.} [Top Quark Working Group Collaboration],
  arXiv:1311.2028 [hep-ph].


\bibitem{Jia:2015uea}
  L.-B.~Jia,
  Phys.\ Rev.\ D {\bf 92}, 074006 (2015)
  [arXiv:1506.05293 [hep-ph]].

\bibitem{D'Hondt:2015jbs}
  J.~D'Hondt, A.~Mariotti, K.~Mawatari, S.~Moortgat, P.~Tziveloglou, G.~Van Onsem,
  arXiv:1511.07463 [hep-ph].


\bibitem{Oh:2009fm}
  For a more careful treatment of light vector boson effects 
  on $B_s$ mixing, see, e.g.
  S.~Oh and J.~Tandean,
  JHEP {\bf 1001}, 022 (2010)
  [arXiv:0910.2969 [hep-ph]].

\bibitem{Ball:2004ye}
  P.~Ball and R.~Zwicky,
  Phys.\ Rev.\ D {\bf 71}, 014015 (2005)
  [hep-ph/0406232].

\bibitem{Ball:2004rg}
  P.~Ball and R.~Zwicky,
  Phys.\ Rev.\ D {\bf 71}, 014029 (2005)
  [hep-ph/0412079].

\bibitem{Beneke:2001at}
  M.~Beneke, T.~Feldmann, D.~Seidel,
  Nucl.\ Phys.\ B {\bf 612}, 25 (2001)
  [hep-ph/0106067].

\bibitem{Williams:2015xfa}
  M.~Williams,
  JINST {\bf 10}, P06002 (2015)
  [arXiv:1503.04767 [hep-ex]].

\bibitem{Aaij:2015tna}
  R.~Aaij {\it et al.} [LHCb Collaboration],
  Phys.\ Rev.\ Lett.\  {\bf 115}, 161802 (2015)
  [arXiv:1508.04094 [hep-ex]].

\bibitem{Hyun:2010an}
  H.J.~Hyun {\it et al.}  [Belle Collaboration],
  Phys.\ Rev.\ Lett.\  {\bf 105}, 091801 (2010)
  [arXiv:1005.1450 [hep-ex]].

\bibitem{Aaij:2012vr}
  R.~Aaij {\it et al.}  [LHCb Collaboration],
  JHEP {\bf 1302}, 105 (2013)
  [arXiv:1209.4284 [hep-ex]].

\bibitem{Aaij:2014pli}
  R.~Aaij {\it et al.}  [LHCb Collaboration],
  JHEP {\bf 1406}, 133 (2014)
  [arXiv:1403.8044 [hep-ex]].

\bibitem{Lees:2013kla}
  J.P.~Lees {\it et al.}  [BaBar Collaboration],
  Phys.\ Rev.\ D {\bf 87}, 112005 (2013)
  [arXiv:1303.7465 [hep-ex]].

\bibitem{Lutz:2013ftz}
  O.~Lutz {\it et al.}  [Belle Collaboration],
  Phys.\ Rev.\ D {\bf 87}, 111103 (2013)
  [arXiv:1303.3719 [hep-ex]].


\bibitem{LHCb:2012de}
  R.~Aaij {\it et al.}  [LHCb Collaboration],
  JHEP {\bf 1212}, 125 (2012)
  [arXiv:1210.2645 [hep-ex]].

\bibitem{Aaij:2015nea}
  R.~Aaij {\it et al.} [LHCb Collaboration],
  JHEP {\bf 1510}, 034 (2015)
  [arXiv:1509.00414 [hep-ex]].


\bibitem{Mescia:2007kn}
  F.~Mescia and C.~Smith,
  Phys.\ Rev.\ D {\bf 76}, 034017 (2007)
  [arXiv:0705.2025 [hep-ph]].


\bibitem{Batley:2011zz}
  J.R.~Batley {\it et al.}  [NA48/2 Collaboration],
  Phys.\ Lett.\ B {\bf 697}, 107 (2011)
  [arXiv:1011.4817 [hep-ex]].

\bibitem{Ecker:1987qi}
  G.~Ecker, A.~Pich, E.~de Rafael,
  Nucl.\ Phys.\ B {\bf 291}, 692 (1987).

\bibitem{AlaviHarati:2000hs}
  A.~Alavi-Harati {\it et al.}  [KTeV Collaboration],
  Phys.\ Rev.\ Lett.\  {\bf 84}, 5279 (2000)
  [hep-ex/0001006].

\bibitem{Mertens:2011ts}
  P.~Mertens and C.~Smith,
  JHEP {\bf 1108}, 069 (2011)
  [arXiv:1103.5992 [hep-ph]].

\bibitem{Batley:2004wg}
  J.R.~Batley {\it et al.}  [NA48/1 Collaboration],
  Phys.\ Lett.\ B {\bf 599}, 197 (2004)
  [hep-ex/0409011].


\bibitem{Artamonov:2008qb}
  A.V.~Artamonov {\it et al.}  [E949 Collaboration],
  Phys.\ Rev.\ Lett.\  {\bf 101}, 191802 (2008)
  [arXiv:0808.2459 [hep-ex]].

\bibitem{Artamonov:2009sz}
  A.V.~Artamonov {\it et al.}  [E949 Collaboration],
  Phys.\ Rev.\ D {\bf 79}, 092004 (2009)
  [arXiv:0903.0030 [hep-ex]].

\bibitem{Artamonov:2005cu}
  A.V.~Artamonov {\it et al.}  [E949 Collaboration],
  Phys.\ Rev.\ D {\bf 72}, 091102 (2005)
  [hep-ex/0506028].

\bibitem{Ahn:2009gb}
  J.K.~Ahn {\it et al.}  [E391a Collaboration],
  Phys.\ Rev.\ D {\bf 81}, 072004 (2010)
  [arXiv:0911.4789 [hep-ex]].

%

\bibitem{Araki:2015mya} 
  T.~Araki, F.~Kaneko, T.~Ota, J.~Sato and T.~Shimomura,
  Phys.\ Rev.\ D {\bf 93}, 013014 (2016)
  [arXiv:1508.07471 [hep-ph]].

\bibitem{Dreiner:2009er}
  H.K.~Dreiner, S.~Grab, D.~Koschade, M.~Kramer, B.~O'Leary, U.~Langenfeld,
  Phys.\ Rev.\ D {\bf 80}, 035018 (2009)
  [arXiv:0905.2051 [hep-ph]].

\bibitem{deVries:2015mfw}
  J.~de Vries, H.K.~Dreiner and D.~Schmeier,
  arXiv:1511.07436 [hep-ph].

\bibitem{NA62}
  See webpage http://na62.web.cern.ch/na62/.

\bibitem{KOTO}
  See webpage http://koto.kek.jp/.

\bibitem{KOTO-CKM2014}
  Talk by K. Shiomi at CKM 2014, Vienna, Austria, September 2014.

\bibitem{LBNE}
  See webpage http://lbne.fnal.gov/.

\bibitem{Aaij:2013iag}
  R.~Aaij {\it et al.} [LHCb Collaboration],
  JHEP {\bf 1308}, 131 (2013)
  [arXiv:1304.6325 [hep-ex]].

\end{thebibliography}
\end{document}